\documentclass[acmsmall,screen,review]{acmart}

\AtBeginDocument{%
  }

\acmConference[Conference acronym 'XX]{Make sure to enter the correct
conference title from your rights confirmation email}{June 03--05,
2018}{Woodstock, NY}

\usepackage[lined, linesnumbered,noend]{algorithm2e}
\usepackage{booktabs}
\usepackage{multicol}
\usepackage{pifont}
\usepackage[noend]{algorithmic}
\usepackage{graphicx}
\usepackage{textcomp}
\usepackage{xcolor}
\usepackage{syntax}
\usepackage{mathtools,bussproofs,turnstile}
\usepackage{listings}
\usepackage{tikz}
\usepackage[switch]{lineno}
\usepackage[inline]{enumitem}
\usepackage[inkscapeformat=png]{svg}
\usepackage{pdfpages}
\usetikzlibrary{shapes, arrows, positioning}
\usepackage[frozencache,cachedir=.]{minted}
\usepackage{tcolorbox}
\usepackage{subcaption}
\usepackage{adjustbox}
\usepackage{svg}
\usepackage{wrapfig}

\floatname{algorithm}{Algorithm}

\newcommand{\mm}[1]{{\textcolor{olive}{(\textbf{Martin:} {#1})}}}

\newcommand{\toolName}{{\sc ProofRover}\xspace}
\long\def\ignore#1{}
\newcommand{\algo}[1]{\textsc{#1}}

\definecolor{deepblue}{rgb}{0,0,0.5}
\definecolor{deepred}{rgb}{0.6,0,0}
\definecolor{deepgreen}{rgb}{0,0.5,0}
\definecolor{halfgray}{gray}{0.55}

\definecolor[named]{ACMBlue}{cmyk}{1,0.1,0,0.1}
\definecolor[named]{ACMYellow}{cmyk}{0,0.16,1,0}
\definecolor[named]{ACMOrange}{cmyk}{0,0.42,1,0.01}
\definecolor[named]{ACMRed}{cmyk}{0,0.90,0.86,0}
\definecolor[named]{ACMLightBlue}{cmyk}{0.49,0.01,0,0}
\definecolor[named]{ACMGreen}{cmyk}{0.20,0,1,0.19}
\definecolor[named]{ACMPurple}{cmyk}{0.55,1,0,0.15}
\definecolor[named]{ACMDarkBlue}{cmyk}{1,0.58,0,0.21}

\definecolor{ckeyword}{HTML}{7F0055}
\definecolor{ccomment}{HTML}{3F7F5F}
\definecolor{cnumber}{HTML}{2A0099}
\definecolor{pblue}{rgb}{0.13,0.13,1}
\definecolor{pgreen}{rgb}{0,0.5,0}
\definecolor{pred}{rgb}{0.9,0,0}
\definecolor{pgrey}{rgb}{0.46,0.45,0.48}

\newcommand{\assertion}[1]{{\footnotesize \textcolor{ACMBlue}{#1}}}

\newcommand{\fail}[1]{#1_{\mathit{fail}}}

\newcommand{\spec}{\varphi}

\newcommand{\pre}{\varphi_\mathit{pre}}
\newcommand{\post}{\varphi_\mathit{post}}

\newcommand{\program}{\mathit{Pr}}
\newcommand{\test}{T}
\newcommand{\nlpintent}{Q}

\newcommand{\vcgen}{\mathtt{VCGen}}
\newcommand{\valid}{\mathtt{Valid}}
\newcommand{\conformsTS}[2]{#1 \subseteq #2}
\newcommand{\conformsPS}[2]{#1 \subseteq #2}

\newcommand{\conforms}[2]{#1 \subseteq #2}

\newcommand{\notconforms}[2]{#1 \nsubseteq #2}

\newcommand{\result}[1]{%
\begin{tcolorbox}[colframe=black,boxrule=0.5pt,arc=4pt,
      left=6pt,right=6pt,top=6pt,bottom=6pt,boxsep=0pt,width=\columnwidth]%
      {#1}
\end{tcolorbox}%
}

\newcommand{\sonnet}{ {\sc Sonnet-3.5} \xspace}
\newcommand{\gptfouro}{ {\sc GPT-4o} \xspace}

\newcommand{\naiverepair}{Naive Repair}
\newcommand{\chainofthought}{Chain-of-Thought}

\begin{document}
\title{Assured Automatic Programming via
Large~Language~Models
}

\author{Martin Mirchev}
\email{mmirchev@comp.nus.edu.sg}
\affiliation{
\institution{National University of Singapore}
\city{Singapore}
\country{Singapore}
}
\author{Andreea Costea}
\email{andre.costea@gmail.com}
\affiliation{%
  \institution{Delft University of Technology}
  \city{Delft}
  \country{The Netherlands}
}

\author{Abhishek Kr Singh}
\email{abhishek.uor@gmail.com}
\affiliation{
\institution{National University of Singapore}
\city{Singapore}
\country{Singapore}
}

\author{Abhik Roychoudhury}
\email{abhik@comp.nus.edu.sg}
\affiliation{
\institution{National University of Singapore}
\city{Singapore}
\country{Singapore}
}

\begin{abstract}
With the advent of AI-based coding engines, it is possible to convert natural language requirements to executable code in standard programming languages. However, AI-generated code can be unreliable, and the natural language requirements driving this code may be ambiguous. In other words, the intent may not be accurately captured in the code generated from AI-coding engines like Copilot. The goal of our work is to \textit{discover the programmer intent},  while generating code which conforms to the intent and a proof of this conformance. 
Our approach to intent discovery is powered by a novel repair engine called program-proof co-evolution, where the object of repair is a tuple (code, logical specification, test) generated by an LLM from the same natural language description. The program and the specification capture the initial operational and declarative description of intent, while the test represents a concrete, albeit partial, understanding of the intent. Our objective is to achieve consistency between the program, the specification, and the test by incrementally refining our understanding of the user intent.
Reaching consistency through this repair process provides us with a formal, logical description of the intent, which is then translated back into natural language for the developer's inspection. The resultant intent description is now unambiguous, though expressed in natural language. We demonstrate how the unambiguous intent discovered through our approach increases the percentage of verifiable auto-generated programs on a recently proposed dataset in the Dafny programming language.
\end{abstract}



\maketitle
\thispagestyle{plain}
\pagestyle{plain}

\section{Introduction}
\label{sec:introduction}

The advent of automatically generated code, such as those produced by Generative AI/Large Language Models (LLMs), offers us new developer workflows.
Tools such as GitHub's Copilot \cite{githubCopilot} have demonstrated the impact automatically generated code may have on productivity \cite{CopilotImpact2024}.
Pre-trained language models perform a wide diversity of programming tasks remarkably well, assuming the right instructions and setup \cite{NashidICSE2023}. 
A typical workflow involves developers describing the desired functionality in a natural language, followed by the AI assistant automatically generating the corresponding code for them. 
Before the integration of the auto-generated code into production, developers are encouraged to check the freshly generated code for safety and correctness \cite{liu2024your}, and potentially ask the AI assistant to regenerate it in case of inconsistencies with the intended output. 
%

This process turns out to be quite laborious in the case where the intent is not met \cite{FanICSEFoSE2023}. 
Assuming a fine-tuned model which performs relatively well, one explanation for the misalignment between intent and output is that natural language descriptions of intent are inherently \textit{ambiguous}, while the output code represents \textit{precise} computations.
In essence, the task of auto-generating code is currently a {one-to-many translation} from natural language to programming language, leading to non-determinism in the software production \cite{ouyang2023llm}. 
Our goal is to streamline this process by removing the ambiguity from the intent -- \textit{discover the intent} -- aiming for a {one-to-one translation} and reducing the non-determinism in the AI-assisted software development. 

Our approach to intent discovery is grounded in software verification, which involves checking that code aligns with the intent as captured in the formal specification. 
We ask the AI assistant to generate both code and its formal specification from the same intent written in natural language. 
At this point the AI assistant generates two transformations, we call them artifacts, in an unambiguous format from the initial ambiguous intent description. 
Next, a verifier checks whether the code meets the formal specification. 
In case the verification fails, that is, the code and its specification do not align, we proceed by extracting the common facts about the intent which the code and its specification both agree upon. 
We use this unambiguous partial intent to start a repair campaign, which we refer to as \textit{program-proof co-evolution}, implemented in a tool we call \toolName.
The objective of this campaign is to repair the code and its specification such that the verification of the repaired program against the repaired specification succeeds.
This results in alignment between the program and its specification, thus ensuring agreement upon the same intent. 
The final repaired specification represents what we infer as the developer's intent written in a formal, unambiguous manner. 
We then translate this formal intent back into natural language for the developer's inspection. 
Although informal, this intent description is now unambiguous. 
The developer can modify this refined intent and restart the code generation process, now with a more precise intent. 

The repair campaign may result in multiple pairs of code and corresponding specifications, thus multiple possible intents. 
To help with the selection of the intent closest to the developer's intent, \toolName supports a third artifact, namely tests. 
Depending on their origin, tests may or may not be trusted to express the developer's true intent. 
If trusted (say, if the developer provides them), tests are used by \toolName to further disambiguate the previously derived common intent, allowing it to filter out code-specification pairs.
If not trusted (say, generated by LLM from the natural language description), tests serve as another source of partial intent,  potentially contributing to the common intent if there is any commonality between the test and the specification or the program. 
%

%
{Generative AI promises to alleviate human burden by automatically synthesizing both the code and its formal specification, given the developer's intent written in natural language.
As it turns out, this promise is not achievable with existing technology as demonstrated in 
a recent empirical study \cite{endres2024can} showing that oftentimes, the auto-generated code and its corresponding auto-generated specifications do not align, thus making the verification fail.
We show how our approach to intent discovery may help the AI assistants produce more verifiable code. }

In summary, we make the following contributions:
\begin{itemize}[topsep=0pt]
    \item We introduce a mechanism for \textit{intent discovery} grounded in formal reasoning. Our core idea is to generate a declarative and an operational description of the intent from LLMs - the specification and the program, respectively. We then extract the common intent from each artifact and use it as the basis for a conformance repair campaign. The campaign repairs \emph{both} the declarative and operational descriptions to ensure their consistency, as confirmed by formal verification. Tests can be used to further clarify the developer's intent.
    
    \item We introduce a novel algorithm named \textit{program-proof co-evolution} to support our idea of developer intent discovery and conformance repair. Specifically, this algorithm distinguishes between a partial, common intent (or \emph{hard} intent) and an unconfirmed intent (or \emph{soft intent}, comprising the facts that are not common across the artifacts). The goal is to resolve the conformance problem by repairing the soft intent while preserving 
    the hard intent.
    
    \item We present \toolName, an implementation of the {program-proof co-evolution} algorithm in the Boogie framework, and present experimental results for Dafny programs. We discuss how the verifiability of code auto-generated from natural language descriptions increases when the intent gets disambiguated. 

\end{itemize}

\section{Overview}
\label{sec:overview}


\subsection{Background}

Owning to its maturity as a programming language that supports software verification via design by contract, we will be using Dafny as our target language throughout this paper \cite{dafnylang}.
However, we believe the ideas presented here can be extended to future verification systems developed to certify auto-generated code.
Dafny programs operate on methods that have declarative specification ascribed to their bodies in the form or pre- and postconditions.
A method is verified using Floyd-Hoare-style verification, where reasoning begins by assuming the ascribed preconditions (specified with the keyword \assertion{\mintinline{C}{requires}}).
The method is then converted to a collection of verification conditions representing the correctness of the method with regards to its specifications, which are sent to a SAT/SMT solver. 
A successful verification of the entire method implies the validity of the ascribed postconditions (specified with the keyword \assertion{\mintinline{C}{ensures}}). 
If the method contains loops, the verifier requires loop annotations to summarize the loop's behavior, even for unbounded executions.

\subsection{Setup}

The repair solution we are proposing for fixing intent
conformance is applicable to a combination 
of human-written and auto-generated code, specifications, and tests.
None of these three artifacts
\begin{wrapfigure}[5]{r}{0.35\linewidth}
\vspace{-1.4em}
\includegraphics[width=\linewidth]{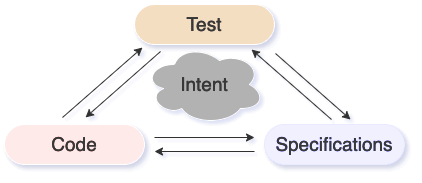}
\end{wrapfigure}
used for expressing intent are assumed correct. They may all represent incorrect or incomplete intent. 
For the simplicity 
of this 
presentation, we are assuming that code and specification are LLM-generated, while tests are provided by an oracle. 
Furthermore, the framework is designed to work with either just two (any two) artifacts or with all three. We will walk through an example using two artifacts, namely code, and specification, and introduce the third, unit tests, to refine the results. 


\subsection{Running example}\label{sec:runningexample}

We ask an LLM to generate a simple Dafny method which finds the first odd value in a given array. 
The LLM generates the imperative code in \autoref{fig:motivating-example-impl}, which appears to be carrying out the expected computation by storing the index of the first odd value in the array in return variable \mintinline{C}{odd}.
The LLM also produces a declarative description for the same intent that, while verbose, appears to be correct: 
%
assuming a non-null array (line \ref{ext:pre}), upon executing the code, \mintinline{C}{odd} stores the index of an odd element (line \ref{ext:postodd}), specifically the first odd element in the array (line \ref{ext:postfirst}). 
The loop invariant (lines \ref{ext:invstart}-\ref{ext:invend}) summarises the effect of the loop\footnote{In Dafny, the variable \mintinline{C}{i} in \mintinline{C}{for i := lo to hi} takes the values from \mintinline{C}{lo} to, \emph{but not including}, \mintinline{C}{hi}.} 
using local variables \mintinline{C}{found} and \mintinline{C}{i} 
to specify the same computation as the method's specification with one notable exception that will be discussed later.

{\bf Verification Results.} 
%
First, the verification fails with index out-of-range errors. As part of its no-runtime-errors safety guarantees, Dafny statically checks that all array accesses are within bounds, a condition that can be proven to hold for the method's body given the loop invariant.
For example, if we were to add the following assertion at the end of \mintinline{C}{FindFirstOdd}, 
Dafny would prove its validity:
{
\centerline{\mintinline[fontsize=\small,autogobble,escapeinside=||,linenos=false]{C}{|\assertion{assert found ==> 0 <= odd < arr.Length;}| }
}}

Additionally, to ensure the well-formedness of the specifications, a similar condition should also hold for them. 
Explicitly, specifications should not quantify over indices that fall outside the
array bounds.
A close inspection of the postconditions (lines \ref{ext:postodd}-\ref{ext:postfirst}) indicates that \mintinline{C}{odd} could refer to an index beyond the array's length. 

Second, the verification fails to prove that
{\mintinline[autogobble,escapeinside=||,linenos=false]{C}{|\assertion{arr[odd]\%2!=0}|}} holds
on all paths, as required  by the 
\begin{wrapfigure}[36]{r}{0.53\linewidth}
\vspace{-1em}
{\begin{center}  
\begin{subfigure}{0.52\textwidth}
\begin{tcolorbox}[size=small,boxrule=0pt,title=Prompt,fonttitle=\footnotesize,]
\footnotesize
\emph{Generate a method that finds the first odd value in an array. Generate the formal specifications which captures this computation and annotate the method with it.}
\end{tcolorbox}
\end{subfigure}
\hspace*{1.5em}
\begin{subfigure}{0.52\textwidth}
\vspace*{1em}
\begin{minted}[fontsize=\small,autogobble,escapeinside=||,numbers=left]{C}
method FindFirstOdd(arr: array<int>) 
            returns(odd: int)
 |\assertion{requires arr != null}| |\label{ext:pre}|
 |\assertion{ensures arr[odd]\,%\,2 != 0}| |\label{ext:postodd}|
 |\assertion{ensures forall i\,::0\,<=\,i\,<\,odd\,==>\,arr[i]\,%\,2\,==\,0 }| |\label{ext:postfirst}|   
{  |\label{ext:start}|
 var found := false;
 odd := -1; 

 for i := 0 to arr.Length
  |\assertion{invariant 0\,<=\,i\,<=\,arr.Length}| |\label{ext:invstart}|
  |\assertion{invariant !found ==> odd == -1}|
  |\assertion{invariant found ==>}||\label{ext:invodd}|
      |\assertion{0\,<=\,odd\,<\,i && arr[odd]\,%\,2\,!=\,0}|
  |\assertion{invariant forall j :: 0\,<=\,j\,<\,i}| |\label{ext:invfirst}| 
      |\assertion{==>((found ==> arr[j]\,%\,2\,!=\,0 $\mid\mid$ j\,==\,odd)}|
     |\assertion{ && (!found ==> arr[j]\,%\,2\,==\,0))}| |\label{ext:invend}|
 {
   if arr[i] % 2 != 0 {
      odd := i;
      found := true;
      break;
   }
 }
}
\end{minted}
\end{subfigure}
\begin{subfigure}{0.52\textwidth}
\vspace*{1em}
\begin{tcolorbox}[size=small,boxrule=0pt,title=Verification Result,width=\textwidth,fonttitle=\footnotesize,]
\footnotesize
line \ref{ext:postodd}: Error 1: index out of range.\\
line \ref{ext:postfirst}: Error 2: index out of range.\\
line \ref{ext:start}: Error 3: A postcondition might not hold on this path.\\
line \ref{ext:postodd}: This is the postcondition that might not hold.\\
No quick fixes available.
\end{tcolorbox}
\end{subfigure}
\end{center}
}
\caption{A Dafny auto-generated method.}
\label{fig:motivating-example-impl}
\end{wrapfigure}
postcondition at line \ref{ext:postodd}. 
However, if added at the end of \mintinline{C}{FindFirstOdd}, the following assertion would be proven valid: \\
\indent{
{\mintinline[autogobble,escapeinside=||,linenos=false]{C}{|\assertion{assert found ==> arr[odd] \%2 != 0;} | }  } }\\
\indent The above assertion seems similar to the invalid postcondition. 
What sets them apart is the presence of the \mintinline{C}{found} guard. 
While simple, this example supports our observation that 
\emph{a well-trained, finetuned LLM generates \underline{almost} correct artifacts.} 
Despite this, turning these artifacts into ones that successfully verify requires non-trivial intervention (suggested by the verification result too: {``No quick fixes available''}).
We next provide an intuition of how \toolName approaches the repair of the conformance problem, indicating how the three errors get fixed. 

\indent {\bf Program-Specification Conformance.}
We first target the program-specification conformance problem indicated by {\small Error 3}.
A program conforms to a specification if the entire set of program behaviors is included in the set of behaviors described by the specification. 
With this interpretation of conformance, \toolName identifies that both the program and specification agree that: \\
\emph{(h1) all elements in the array up to the index denoted by \mintinline{C}{odd} should be even numbers. }

It also identifies that the program and specification do not agree on the followings: \\
(s1) \emph{when \mintinline[linenos=false]{C}{found} is true,  \mintinline[linenos=false]{C}{odd} indexes into an odd element in \mintinline[linenos=false]{C}{arr}.}\\
(s2) \emph{\mintinline[linenos=false]{C}{odd} always indexes into an odd element in \mintinline[linenos=false]{C}{arr}.}

The common intent described in statement (h1) is supported by the fact that Dafny can prove that the postcondition at line \ref{ext:postfirst} is implied by the program path on which the invariants at lines \ref{ext:invstart} and \ref{ext:invfirst} hold. 
The intent described by statement (s1) and extracted from the program is supported by the invariant at line \ref{ext:invodd}. The intent described by statement (s2) and extracted from the specification is supported by the postcondition at line \ref{ext:postodd}. 
To simplify the explanation of this example, we described the intent expressed by statements (h1), (s1), and (s2) using natural language; however, \toolName extracts logical formulas from Dafny code to represent these facts of the intent. The definition of facts and the detailed mechanics for deriving them are deferred to \autoref{sec:method}.

After extracting all the facts, \toolName then differentiates them into \emph{hard intent} and \emph{soft intent} before initiating the repair. 
The hard intent is a collection of facts which should still hold after the repair. The common intent described by (h1) and the well-formedness statement (h2) \emph{all array accesses should be within bounds} are examples of facts that represent hard intent.
The soft intent is a collection of facts which need not hold after the repair. Practically, the soft intent contains all the facts that are not part of the hard intent. Conceptually, the soft intent represents facts on which the various artifacts do not reach full agreement.
For instance, the formulas supporting statements (s1) and (s2) are examples of soft intent facts.

The goal of the repair is to modify the soft intent such that all artifacts respect their conformance relation with each other while maintaining the validity of the hard intent. At the same time, it seeks to preserve as many soft constraints as possible, aiming for a less intrusive repair.

\begin{figure}[t!]
\centering
\begin{subfigure}{0.32\textwidth}
\begin{minted}[fontsize=\small,autogobble,escapeinside=||,numbers=none]{C}
// |@|trust
method OddInArray()
{
 var x := new int[]{2,3,4};
 var s := FindFirstOdd(x);
 |\assertion{assert s >= 0;}|
}
\end{minted}
\caption{}\label{fig:motivating-example-impl-simple-test}
\end{subfigure}
\hfill
\begin{subfigure}{0.32\textwidth}
\begin{minted}[fontsize=\small,autogobble,escapeinside=||,numbers=none]{C}
// |@|trust
method AllEven()
{
 var x := new int[]{2,2,4};
 var s := FindFirstOdd(x);
 |\assertion{assert s == -1;}|
}
\end{minted}
\caption{}\label{fig:motivating-example-impl-alleven}
\end{subfigure}
\hfill
\begin{subfigure}{0.32\textwidth}
\begin{minted}[fontsize=\small,autogobble,escapeinside=||,numbers=none]{C}
// |@|trust
method AllEvenLength()
{
 var x := new int[]{2,2,4};
 var s := FindFirstOdd(x);
 |\assertion{assert s == -x.Length;}|
}
\end{minted}
\caption{}\label{fig:motivating-example-impl-alleven-length}
\end{subfigure}
\caption{Small test case for \mintinline{C}{FindFirstOdd}.}
\Description{Small test case from the user to present new behavior. }
\end{figure}

For {\small Error 3}, \toolName identifies that it is the misalignment of statements (s1) and (s2) that lead to the conformance issue. 
It could then select statement (s1), statement (s2), or both statements as repair candidates.
Choosing statement (s2) as the repair candidate seems to only involve the refinement of the postcondition at line \ref{ext:postodd}. Weakening it as indicated below
makes {\small Error 3} disappear without invalidating any of the hard constraints:

{\setlength\intextsep{2pt}
\begin{figure}[h]
\begin{minipage}[t]{0.6\textwidth}
\centering
    \renewcommand\theFancyVerbLine{%
        \ifnum\value{FancyVerbLine}=1
            \tiny\setcounter{FancyVerbLine}{4}4
        \else
            \ifnum\value{FancyVerbLine}=5
            \tiny\setcounter{FancyVerbLine}{4}4
        \else
            {\tiny\arabic{FancyVerbLine}}%
        \fi
        \fi
}
\begin{minted}[fontsize=\small,autogobble,escapeinside=||,numbers=left,xleftmargin=4.0ex]{C}
 ---  |\assertion{ ensures arr[odd] \%2 != 0;}|
 +++  |\assertion{ ensures  0 <= odd < arr.Length ==> arr[odd] \%2 != 0;}|
\end{minted}
\end{minipage}
\Description{Patch ensuring well-formedness of the first postcondition.}
\end{figure}
}

Later in this section, we discuss how to automatically prioritize which soft constraints to repair. 

Upon examining the reasons for the proof failures behind {\small Error 1} and {\small Error 2}, it becomes obvious that the issues are related solely to the specification's conformance to the well-formedness condition rather than to program-specification conformance. We ignore {\small Error 1} since the fix for {\small Error 3} is sufficient to also fix this failure. In fixing {\small Error 2}, \toolName identifies the postcondition at line \ref{ext:postfirst} as a fact that is at odds with the well-formedness condition, which is part of the hard intent. Aiming to align the two while maintaining the hard intent leads to the following fix:

{\setlength\intextsep{2pt}
\begin{figure}[h]
\begin{minipage}[t]{0.9\textwidth}
\centering
    \renewcommand\theFancyVerbLine{%
        \ifnum\value{FancyVerbLine}=1
            \tiny\setcounter{FancyVerbLine}{5}5
        \else
            \ifnum\value{FancyVerbLine}=6
            \tiny\setcounter{FancyVerbLine}{5}5
        \else
            {\tiny\arabic{FancyVerbLine}}%
        \fi
        \fi
}
\begin{minted}[fontsize=\small,autogobble,escapeinside=||,numbers=left,xleftmargin=4.0ex]{C}
 ---  |\assertion{ensures forall i::0 <= i < odd ==> arr[i] \% 2 == 0; }|
 +++  |\assertion{ensures  0 <= odd < arr.Length ==> (forall i::0 <= i < odd ==> arr[i] \% 2 == 0);}|
\end{minted}
\end{minipage}
\Description{Patch ensuring well-formedness of the second postcondition. }
\end{figure}
}

Details about how \toolName manipulates the soft and hard intents to find a fix are presented in \autoref{sec:method}. There, we also dicuss their formal definition.  

{\bf Specification-Test Alignment.} One might argue that the specification fixes that \toolName found are not unique. 
Na\"{i}ve fixes, such as replacing both postconditions from \autoref{fig:motivating-example-impl} with 
{
{\mintinline[autogobble,escapeinside=||,linenos=false]{C}{|\assertion{ensures true; }|}}}
-- an obviously vacuous solution --
would also fix the conformance problem. 
While a human would not choose to repair the method this way, it is not clear how to instruct a repair engine to avoid such vacuous solutions if the goal of the repair were to simply make the verification succeed.
Luckily, our repair framework lends itself well as a filter mechanism for vacuous specifications when unit tests are present. 
The reason for that is that the existence of a unit test, e.g., such as the one in \autoref{fig:motivating-example-impl-simple-test}, is still a verification problem for Dafny. 
The call to the function being tested which is described by its formal specification, must conform to the test constraints: given an input array 
{{\mintinline[autogobble,escapeinside=||,linenos=false]{C}{int[]{2,3,4}}}}
the result returned by {{\mintinline[autogobble]{C}{FindFirstOdd}}} should be greater or equal to zero. 
In other words, the conformance of specification to the test implies that the behavior described by the specification \emph{in the considered calling context} is a subset of the behaviors captured by the unit test. 
Furthermore, if we could specify that this test is part of the hard intent\footnote{\toolName supports user annotations to indicate that an artifact is trusted, i.e., it is part of the hard intent.}, then \toolName could use this information to restrict the search space of specification fixes accordingly and thus eliminate the spurious fix. 
In the case of the spurious specification, \toolName detects the verification failure of 
{{\mintinline[autogobble,escapeinside=||,linenos=false]{C}{|\assertion{assert s >= 0}|}}}
in \autoref{fig:motivating-example-impl-simple-test} and uses this failure information to restrict the conditions over the return value {{\mintinline[autogobble]{C}{odd}}} in 
{{\mintinline[autogobble]{C}{OddInArray}}}.
The spurious postcondition {{\mintinline[autogobble]{C}{true}}} fails to imply {{\mintinline[autogobble,escapeinside=||,linenos=false]{C}{|\assertion{odd >= 0}|}}} for the given input array, and it is thus considered a non-solution. 

\begin{figure}[t]
    \centering
    \includegraphics[width=\linewidth]{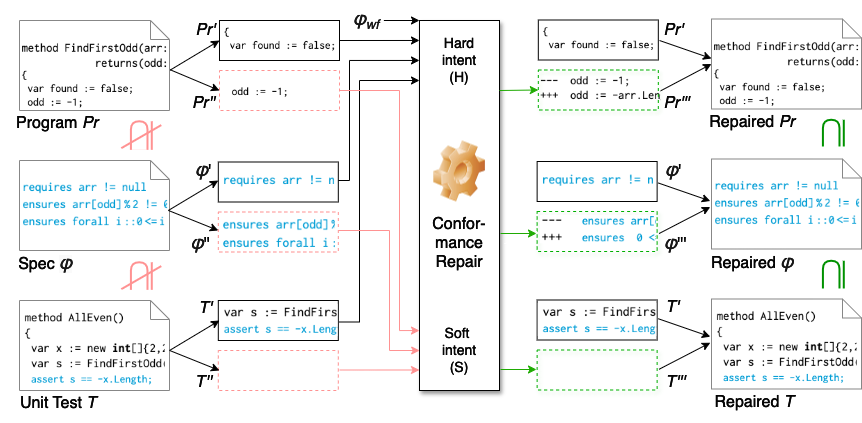}
    \caption{The manipulation of soft and hard intents for repairing the conformance issue in \autoref{fig:motivating-example-impl}.}
    \label{fig:overview}
\Description{The manipulation of soft and hard intents for repairing the conformance issue in the motivating example.}
\end{figure}

Let us next consider a unit test with an input array that contains no odd element, like the one in \autoref{fig:motivating-example-impl-alleven}, also marked as part of the hard intent. 
Although \mintinline{C}{FindFirstOdd} does return -1 in such a calling context, its ascribed specification does not describe this scenario, deeming the assertion    {{\mintinline[autogobble,escapeinside=||]{C}{|\assertion{s == -1}|}}} as invalid.
This failure triggers the repair engine which subsequently discovers that a new postcondition should be ascribed to \mintinline{C}{FindFirstOdd}, namely:

{\setlength\intextsep{3pt}
\begin{figure}[h]
\begin{minipage}[t]{0.9\textwidth}
\centering
    \renewcommand\theFancyVerbLine{%
        \ifnum\value{FancyVerbLine}=1
            \tiny\setcounter{FancyVerbLine}{6}6
        \else
            {\tiny\arabic{FancyVerbLine}}%
        \fi
}
\begin{minted}[fontsize=\small,autogobble,escapeinside=||,numbers=left,xleftmargin=4.0ex]{C}
 +++  |\assertion{ensures (forall i :: 0 <= i< arr.Length ==> arr[i] \% 2 == 0) ==> odd == -1}|
\end{minted}
\end{minipage}
\Description{Example patch for inferring a new specification.}
\end{figure}
}

\noindent thus leading to a more ``complete'' specification. 

Finally, let us consider the unit test in \autoref{fig:motivating-example-impl-alleven-length}, which if marked as part of the hard intent imposes repair obligations not only in the specification, but also, transitively, at the program level too: 

{\setlength\intextsep{3pt}
\begin{figure}[h]
\begin{minipage}[t]{0.9\textwidth}
\centering
    \renewcommand\theFancyVerbLine{%
        \ifnum\value{FancyVerbLine}=1
            \tiny\setcounter{FancyVerbLine}{6}6
        \else
            \ifnum\value{FancyVerbLine}=7
            \tiny\setcounter{FancyVerbLine}{8}8
        \else
            \ifnum\value{FancyVerbLine}=9
            \tiny\setcounter{FancyVerbLine}{8}8
        \else
            {\tiny\arabic{FancyVerbLine}}%
        \fi
        \fi
        \fi
}
\begin{minted}[fontsize=\small,autogobble,escapeinside=||,numbers=left,xleftmargin=4.0ex]{C}
 +++  |\assertion{ensures (forall i :: 0 <= i< arr.Length ==> arr[i] \% 2 == 0) ==> odd == -arr.Length}|
 ---  odd := -1;
 +++  odd := -arr.Length;
\end{minted}
\end{minipage}
\Description{Example patch for aligning with new user intent.}
\end{figure}
}

{\bf Prioritising the repair of the soft intent facts.}
Up to this point, we have not differentiated between the facts of a soft intent; 
\toolName appears to have effectively selected the fact to be resolved that resulted in the least intrusive fix for program-specification conformance.
To come to this result, we have actually devised a few heuristics for choosing a candidate fact to be fixed. 
For example, when choosing to repair (s2) instead of (s1) earlier, the rationale behind that choice was that (s2) seems to be at odds with (h2), the well-formedness constraint, while (s1) is not. 
Generally, we choose the fact that is inconsistent with the most hard intent facts. 
If two facts break even, we next choose the one which is inconsistent with the soft intent facts. 
If they further break even, we then choose the strongest among the two if they are in an implication relation, i.e. (s2) {{\mintinline[autogobble]{C}{==>}}} (s1), or pick randomly otherwise. 
For the unit test in \autoref{fig:motivating-example-impl-alleven}, even if the test were not trusted, that is, it would be considered a soft intent, \toolName would still produce the same patch. That is because the postcondition would be inconsistent with both the program and the unit test, which both indicate that the returned index should be -1 for an array containing only even elements.

{\bf Putting it all together.}
\autoref{fig:overview} summarises our discussion so far. Given at least two of the following artifacts: program ($\program$), specification ($\spec$), and unit test ($\test$), \toolName aims to ensure that any pair of these artifacts is in a conformance relation (sometimes transitively). To do this, \toolName extracts all the facts from the three artifacts and divides them into facts common across all artifacts (the set of which forms the common intent) and facts which are inconsistent across any two artifacts (some of which are responsible for breaking the conformance relation). The facts in the common intent together with the well-formedness conditions ($\spec_\mathit{wf}$) form the hard intent. The remaining facts form the soft intent. The purpose of the repair is to refine the soft intent, preserving the hard intent. 
The outcome consists of patches for the soft intent that, when applied to their corresponding artifacts, ensure a conformance relation holds between each pair of artifacts.
Facts are logical abstractions that \toolName extracts from each artifact using Boogie, Dafny's verification system. Additionally, Boogie's capability to partition a program into slices of proof obligations (assumptions and assertions) allows \toolName to identify which pairs of facts are expected to be in a conformance relation \cite{dafnyrefmanual}. 
How \toolName extracts some of these logical abstractions is described in \autoref{sec:background}, 
while the co-evolution process is introduced in \autoref{sec:method}
%

\section{Background: Dafny - A Verification Aware Programming Language}\label{sec:background}
Dafny \cite{dafnylang} is a strongly typed imperative programming language that supports Hoare-style code verification \cite{hoareAxiomatic}. 
A Dafny program $\program$ consists of modules containing classes, methods, functions, and lemmas that can be annotated with specifications $\spec$, pairs of formulas in the first-order logic. We denote these formulas as $\pre$ and $\post$, respectively, representing the assumed precondition and ascribed postconditions to the method.
To verify a program's correctness, Dafny translates the annotated code into Boogie Intermediate Language methods, representing verification requirements, such as method correctness and signature well-formedness. 
Method correctness ensures that, given a precondition $\pre$, the post-state satisfies the postcondition $\post$.
It also verifies the well-formedness of statements $\spec_\mathit{wf}$ and intermediate assertions $\spec_\mathit{im}$. 
Signature well-formedness checks that pre- and postconditions $\pre, \post$ do not allow ill-formed models, such as out-of-bounds errors (see \autoref{sec:overview}).

{\bf Generating Verification Conditions.} 
The Boogie tool constructs a control flow graph for each method, described by a small programming language~\cite{barnett2005weakest-precondition} (see \autoref{fig:verificationLang}
). 
The transformation of a Boogie method into this graph is called program passification.
This passive program is tailored towards verification by focusing on logical relationships and eliminating side effects. 
The graph is then converted into logical equations using weakest precondition calculus \cite{barnett2005weakest-precondition}, forming the verification conditions. 
The process of converting a Dafny program $\program$ and specification $\spec$ into verification conditions is covered by the function $\vcgen$ whose details are skipped for brevity \cite{barnett2005weakest-precondition}. 
As an example, for the program 
{{\mintinline[autogobble,escapeinside=||,linenos=false]{C}{var x:=1; var y:=2;}}}
and specification 
({{\mintinline[autogobble,escapeinside=||,linenos=false]{C}{|\assertion{assume true;}|}}},
{{\mintinline[autogobble,escapeinside=||,linenos=false]{C}{|\assertion{assert x + y >= 2}|}}})
$\vcgen$ produces the following formula $\mathtt{true} \Rightarrow ((x = 1) \Rightarrow ((y = 2) \Rightarrow (x + y \geq 2)))$.

\begin{figure}[t]
    \footnotesize
    \centering
     \begin{subfigure}[b]{0.7\textwidth}
        \begin{grammar}
            <program> ::= <block>+  \qquad \qquad <block> ::= <ident> `:' <stmt> `;' `goto' <blockId>*
            
            <stmt> ::= `assert' <expr>
            $\mid$ `assume' <expr>
            $\mid$ <stmt> `;' <stmt>
            $\mid$ `skip'        
        \end{grammar} 
    \end{subfigure}
    \caption{The Verification Language}
    \label{fig:verificationLang}
\Description{Verification language used in Boogie.}
\end{figure}    

{\bf Verification with Z3.}
After generating the verification conditions, Boogie interfaces with the Z3 theorem prover for validity check.
We will use a function $\valid$ to denote the validity check on verification conditions. 
The results obtained from Z3 can fall into two main categories: \emph{proof of correctness} or \emph{proof failure}. 
Suppose Z3 proves that the verification conditions are valid. 
In that case, it essentially means that, for all possible models the verification conditions are satisfied. 
This can also be viewed as a conformance relation $\conformsPS{\program}{\spec}$ between a program $\program$ and specification $\spec$:

\centerline{ $ \conformsPS{\program}{\spec} \triangleq \valid(\vcgen(\program,\spec)) $}

If Z3 cannot prove the verification conditions to be valid, it indicates that at least one scenario exists in which the properties do not hold or the solver was unable to prove the target expression due to incompleteness.
Boogie collects the verification result by Z3. 
This information is crucial for debugging and subsequently repairing the program.
Using this proof failure we obtain a failing execution trace $t_{\mathit{fail}} = s_1. s_2 \ldots s_f$ of blocks and statements from the control flow graph that highlights a portion of the Boogie code where the failure occurred. This Boogie fragment can be mapped back to a fragment from the original Dafny program $\program$.

{\bf Tests in Dafny.}
We define a test $\test$ as a pair $(I, O)$ of an input variable $I$ and a logical formula $O$. 
Tests in Dafny can be static or dynamic. 
In this work, we will focus on static tests.

To check for conformance of a test $\test$ with a program $\program$, we can construct a specification $\spec_{T} = (\spec^{I}_{pre}, \spec^{O}_{post})$ that acts as a specification for a program $\program$, allowing for direct verification of the implementation code. We define the relation $\conformsTS{\program}{\test}$ to check for such conformance. 

\centerline{ $\conformsTS{\program}{\test} \triangleq \conformsPS{\program}{\spec_{T}}$}

To check for conformance between a specification $\spec$, say, of method  \mintinline{C}{FindFirstOdd}, and a test $\test$, say, \mintinline{C}{OddInArray}, we construct a program $\program_{T}$ whose specification constrains the parameters and return value of $\program_{T}$ according to $(I, O)$.   
We present an example construction in \autoref{fig:example-test-spec-conformance}
Using this construct, we can define conformance relation $\conformsTS{\spec}{\test}$ between a specification $\spec$ and a test $\test$:

\begin{figure}
\centering
\begin{subfigure}{0.4\textwidth}
\begin{minted}[fontsize=\footnotesize,autogobble,escapeinside=||,numbers=none]{C}
method OddInArray()
{
 var x := new int[]{2,3,4};  // |$I$|
 var s := FindFirstOdd(x);
 |\assertion{assert s >= 0;}|               // |$O$|
}
\end{minted}
\end{subfigure}
\hfill
\begin{subfigure}{0.5\textwidth}
\begin{minted}[fontsize=\footnotesize,autogobble,escapeinside=||,numbers=none]{C}
method OddInArray(int[] x) returns (s:int)
  |\assertion{requires x == new int[]\{2,3,4\}}|    |$  \spec^{I}_{pre} $|
  |\assertion{ensures s >= 0}|                  |$\spec^{O}_{post} $|
{
 s := FindFirstOdd(x);
}
\end{minted}
\end{subfigure}
\caption{Unit test translation from original $(I,O)$ pair (left) to program $\program_T$ and specification $\spec_T$ (right).}
\label{fig:example-test-spec-conformance}
\Description{Unit test translation of test to program and specification.}
\end{figure}

\centerline{ $\conformsTS{\spec}{\test} \triangleq  \conformsPS{\program_{\spec}}{\spec_{\test}} $}

This approach to conformance checking allows us to treat conformance in a uniform manner across all artifacts, regardless which pair of artifacts is considered for repair. 

{\bf Assertion Partitioning.}
To allow for multiple assertions along a path and ease the proving for the underlying theorem prover, Boogie applies partitioning over the generated assertions \cite{dafnyrefmanual}. This allows a user to examine multiple failing conditions instead of stopping the verification process upon encountering the first verification failure. For example, a simple assertion such as {\small $\mathtt{assert} \ \spec_{1} \land \spec_{2}$ } is decomposed into separate assertions {\small $\mathtt{assert} \ \spec_{1}; \mathtt{assert} \ \spec_{2}$}. Furthermore, Boogie constructs unique control flow graph paths, allowing to separately validate these verification conditions.
This partitioning, combined with the failing execution trace, enables a more granular view of the conformance between a program $\program$ and its specification $\spec$. 
Instead of treating the entire program and specification as a monolithic pair, it decomposes the overall conformance problem into a collection of smaller, more manageable pairs, each consisting of a sub-program $\program'$ and a corresponding sub-specification $\spec'$. 
This decomposition allows for targeted repair efforts, where each pair $(\program', \spec')$ can be repaired independently.
This partitioning facilitates the extraction of the soft and hard intent, which will be discussed in \autoref{sec:method}.

\section{Program Proof Co-Evolution}
\label{sec:method}







In this section, we detail how \toolName is designed to discover and formalize the programmer's true intent with the overarching goal of facilitating formally verified auto-generated programs.
Our workflow, depicted in \autoref{fig:coevolution} and named \emph{program-proof co-evolution},  involves three key entities: the programmer, a large language model (LLM), and a verification engine. 
The programmer initiates the workflow by prompting the LLM with a natural language query $\nlpintent$ that conveys their intent. The LLM generates a Dafny program $\program_0$ and corresponding specification $\spec_0$.
A program annotated this way is sent to Dafny for verification (nodes \textbf{2} and \textbf{3} in \autoref{fig:coevolution}).
If the verification fails, that is, $\notconforms{\program}{\spec}$, then \toolName first identifies the hard and soft intents (nodes \textbf{4}) using the partitioning described in \autoref{sec:background}, and then proceeds to generate patches for the soft intent (nodes \textbf{5}).
These patches are applied to $(\program,\spec)$, the original pair of program and specification, before starting the verification process again (nodes \textbf{2}-\textbf{5}) to check whether further refinement is required.

Additionally, \toolName also considers unit tests represented as $\test$ (node \textbf{1}).
However, each iteration of the co-evolution process (nodes \textbf{2}-\textbf{5}) focuses on the conformance of exactly two artifacts, such as the program and specification.
A third artifact, like the test $\test$, is only integrated in a subsequent iteration of the co-evolution, along with the outcome of the previous iteration, for instance, the repaired program-specification conformance.
At a high level, for a triple $(\program, \spec, \test)$ to be accepted, our workflow aims to solve the following two  problems in an interleaved way: 

\begin{enumerate}
    \item[\bf P1] Conformance between a program $\program$ and specification $\spec$, i.e. $\conformsPS{\program}{\spec}$.
    \item[\bf P2] Refining the conforming solutions, as conveyed through feedback from tests. 
\end{enumerate}

Solving the conformance problem {\bf P1} is achieved through a cyclic and automated interaction between the co-evolution driver and the verification engine (nodes \textbf{2}-\textbf{5}).
Through iterative cycles, the co-evolution driver refines the annotated program and its specification based on feedback from the verification engine.
This automated process minimizes the manual interventions needed from the programmer, allowing for a more efficient workflow.  

Refining the conforming solutions, as outlined in problem {\bf P2}, is achieved via a carefully designed interaction between the programmer and the co-evolution process (nodes \textbf{6}-\textbf{8}). This interaction crystalizes the programmer's true intent, justifying the treatment of the test set $T$ as \emph{hard intent} in our running example in \autoref{sec:runningexample}.  We now elaborate on each of these essential components.

\begin{figure*}
\centering
\resizebox{0.6\columnwidth}{0.3\textheight}{
    \begin{tikzpicture}[node distance=1cm and 2cm, auto, font = \small]
    
        \tikzstyle{block} = [rectangle, draw, fill=blue!10, text width=5em, text centered, rounded corners, minimum height=2em]
        \tikzstyle{blockG} = [rectangle, draw, fill=green!10, text width=5em, text centered, rounded corners, minimum height=2em] 
        \tikzstyle{cloudB} = [cloud, draw, fill=blue!20, text width=5em, text centered, aspect=2 ]

        \tikzstyle{txtblkg} = [rectangle, fill=green!5, text width=5em, text centered, rounded corners, minimum height=2em]
        \tikzstyle{txtblkb} = [rectangle, fill=blue!5, text width=5em, text centered, rounded corners, minimum height=2em]
        \tikzstyle{txtblkr} = [rectangle, fill=red!5, text width=5em, text centered, rounded corners, minimum height=2em]

        \tikzstyle{circleG} = [circle, draw, fill=green!20, text width=5em, text centered]
        
        \tikzstyle{line} = [draw, -latex']
    
        \node [block, label={195:{\textbf{2}.}}] (dafny) {Annotated Program $\program$, $\spec$};
        \node [block, above=of dafny,  fill=blue!20, yshift=0.2cm, label={183:{\textbf{1}.}}] (pst)  {$\program_{0}, \spec_{0}, \test_0$};

        \node [block, above=of pst,  fill=blue!20, label={195:{\textbf{0}.}}] (nl)  {Natural Language $Q$ \\ (START)};
        \node [block, right=of dafny, xshift=0cm, label={195:{\textbf{3}.}}] (boogie) {Verification Engine (Boogie)};
        
        \node [block, below=of boogie, fill=red!10, xshift=0cm, yshift=0cm, label={200:{\textbf{4}.}}] (trace) {Hard/Soft Intents \\ $\phi_H$, $\phi_S$};
        \node [txtblkr, right=of trace, xshift=-1.7cm] (notmodels) {$\notconforms{Pr}{\varphi}$};

        
        \node [block, right=of boogie, xshift=0.0cm,label={200:{\textbf{6}.}}] (aap) 
        {Test T};
        
        \node [txtblkb, below=of aap, yshift=0.9cm] (models) {$\conforms{Pr}{\varphi}$};
        
        \node [cloudB, above=of aap, xshift=0.0cm, yshift=0cm, cloud puffs=20, label={250:{\textbf{7}.}}] (dev) {Programmer's Intent};
        
        \node [blockG, above=of dev, fill=green!20, label={190:{\textbf{9}.}}] (stop) {Accepted (STOP)};
        \node [txtblkg, left=of stop, xshift=1.5cm] (alsomodels) {$\program \subseteq \spec \subseteq \test $};

        \node [block, left=of dev, xshift=0.5cm, fill=red!10, label={270:{\textbf{8}.}}] (feedback) {Test $T'$};

        \node [block, left=of trace, fill=blue!20, label={185:{\textbf{5}.}}] (evolve) {Candidate synthesis};

        \path [line] (nl) -- node {LLM} (pst);
        \path [line] (pst) -- node {} (dafny);
        \path [line] (dafny) -- node {$\vcgen$} (boogie);
        \path [line] (boogie) -- node {Fail} (trace);
        
        \path [line, dashed] (boogie) -- node {Pass} (aap);
        \path [line] (aap) -- node {} (dev);        
        \path [line] (dev) -- node {Agree} (stop);
        \path [line] (dev) -- node {Refine} (feedback);
        \path [line, dashed] (feedback) -- node {Feedback} (pst);

        \path [line] (trace) -- node {} (evolve);
        \path [line] (evolve) -- node { $\program''',\spec'''$ } (dafny);
     
    \end{tikzpicture}  }
  \caption{Programmer's intent discovery through Program Proof co-evolution, where the programmer bootstraps the whole process at node \textbf{1}. with a natural language query, and the co-evolution process finally terminates at node \textbf{11}. with the alignment of the programmer's true intent and the actual behavior of the repaired Program.}
  \label{fig:coevolution}
  \Description{Diagram of Program-Proof co-evolution}
\end{figure*}

\subsection{P1: Conformance of Program and Specification}

To verify that $\conformsPS{\program_{0}}{\spec_{0}}$ where $\spec_{0}=(\pre,\post)$, the pair $(\program_{0}, \spec_{0})$ is converted by $\vcgen$ into a logical expression, which is then checked for validity using an SMT solver (refer to node \textbf{3}). 
For now, we ignore the existence of tests.
Suppose the outcome of the validity check is successful. In that case, it confirms that $\conformsPS{\program_{0}}{\spec_{0}}$, allowing us to proceed with a series of steps to clarify the programmer's intent through a controlled interaction (detailed in \autoref{subsec: AlignIntent}). 

{\bf Extracting Hard and Soft Intent.}
Suppose the validity check is unsuccessful. The verifier returns a failing trace of blocks of statements $t_{\mathit{fail}} = s_1. s_2 \ldots s_f$.
The trace $s_{1}. s_{2} \ldots s_{f-1}$ corresponds to a subprogram $\fail{I}$ from $\program_{0}$. 
The final statement $s_{f}$ corresponds to an unproven assertion $\spec_{\mathit{fail}}$  which may come from the postcondition $\post$, intermediate assertions $\spec_{im}$ or well-formedness checks $\spec_{wf}$.
A pair so formed, namely, ($I_{\mathit{fail}},(\pre,\spec_{\mathit{fail}})$), represents a partition of the larger pair ($\program_{0},\spec_{0}$), where $\notconforms{I_{\mathit{fail}}}{(\pre, \spec_{\mathit{fail}})}$. 
To fix this nonconformance, we must first define the space of statements over which patches can be constructed. 
We need to discover both the \emph{commonality} of intent, which is represented by all conforming partitions $(\program', \spec')$ and the \emph{discrepencies} in intent, which is represented by all nonconforming partitions $(\program'', \spec'')$ . 
We denote the collection of verification conditions corresponding to all the pairs $(\program', \spec')$ as the hard intent $\phi_{H}$ and the collection of verification conditions corresponding to all pairs $(\program'',\spec'')$ as the soft intent $\phi_{S}$.
We detail the identification of the {hard intent} $\phi_H$ and the {soft intent} $\phi_S$ in \autoref{alg:intent-extraction}, $\algo{ExtractHSIntent}$.


\RestyleAlgo{ruled}
\begin{wrapfigure}[18]{r}{0.52\linewidth}
\vspace{-1.5em}
\begin{minipage}{0.52\textwidth}
\begin{algorithm}[H]
\caption{$\algo{ExtractHSIntent}$}
\label{alg:intent-extraction}
\small
\DontPrintSemicolon
\LinesNumbered
\SetNoFillComment
\SetFuncSty{textsc}

\SetKw{Skip}{skip}
\SetKw{Break}{break}
\SetKw{Continue}{continue}

\SetKwFunction{HasCandidates}{HasCandidates}
\SetKwFunction{ExtractCandidate}{ExtractCandidate}
\SetKwFunction{IsTrusted}{IsTrusted}
\SetKwFunction{IsWF}{IsWF}
\SetKwFunction{TransformWF}{TransformWF}
\SetKwFunction{ExtractDafnyCode}{ExtractDafnyCode}
\SetKwFunction{PromptLLMForCandidate}{PromptLLMForCandidate}

\KwIn{Program $\program$, Spec $\spec$}
\KwOut{Hard intent $\phi_{H}$, Soft intent $\phi_{S}$}
$\mathcal{P} \gets $ retrieve all partitions of $(\program,\spec)$;

$\phi_{H} \gets \{ \}; \quad \phi_{S} \gets \{ \};$

\For{ ${(\program_i,\spec_i)} \in \mathcal{P}$ }
{
    \uIf{ $\conformsPS{\program_i}{\spec_i}$}
    {
    $\phi_{H} \gets \phi_{H} \cup \vcgen({\program_i},{\spec_i});$
    }
    \uElse
    {
        \For{ $s_t \in \vcgen(\program_i,\spec_i) $ }
        {
            \uIf{$\IsTrusted(s_t)$}{
         
                $\phi_{H} \gets \phi_{H} \cup \{s_t\};$
                
                \Continue;
            }
            \uIf{$\IsWF(s_t)$}{
             
                $\phi_{H} \gets \phi_{H} \cup {\TransformWF(s_t)};$
                
                \Continue;
            }
            
            $\phi_{S} \gets \phi_{S} \cup \{s_t\};$
        }
    }
}

$\phi_{S} \gets $ remove facts from $\phi_{S}$ that are also in $\phi_{H};$
\end{algorithm}
\end{minipage}
\end{wrapfigure}

\autoref{alg:intent-extraction} examines all the partitions (line 1) that come from $(\program,\spec)$. All formulae of the conforming pairs are included in the hard intent (lines 4 and 5). 
If the pair is nonconforming, we must examine each formula $s_t$ of the failing pair (line 7). 
Suppose that $s_{t}$ originates from a statement that is annotated\footnote{This annotation is represented by an attribute {\mintinline{C}{:trusted}} in the Dafny code, which can be attached to any statement.} 
 in the original program or specification as a hard intent by the user (checked using a method called $\mathtt{IsTrusted}$ - line 8).
In that case, we add it as a fact of the hard intent (lines 9). 
Otherwise, we check whether it comes from a well-formedness check (checked using a method called $\mathtt{IsWF}$).
Well-formedness constraints $\spec_{wf}$ represent the underlying safety conditions of the language. 
We must track them as a fact of the hard intent with a single caveat - we do not take the original formula but a modified version that tracks whether the statement for which this check was created exists in the program. 
We represent this aforementioned statement transformation by a method $\mathtt{TransformWF}$. If the formula does not fall into these two cases, we include it as a fact of the soft intent (line 14).
Finally, we ensure that both collections are disjoint, where a formula $s_{i} \in \phi_{S} \cap \phi_{H}$ will be kept in the hard intent (line 15).

{\bf Synthesizing Patches.} Having collected the soft intent $\phi_{S}$ and hard intent $\phi_{H}$, we have to synthesize a patch 
which fixes the $\notconforms{I_\mathit{fail}}{(\pre,\spec_\mathit{fail})}$ nonconformance.
We achieve this using \autoref{alg:synthesis}, \RestyleAlgo{ruled}
\begin{wrapfigure}[11]{r}{0.55\linewidth}
\vspace{-1.5em}
\setlength{\algomargin}{8pt}
\begin{minipage}{0.55\textwidth}
\begin{algorithm}[H]
\caption{$\algo{SynthesizePatch}$}
\label{alg:synthesis}
\small
\DontPrintSemicolon
\LinesNumbered
\SetNoFillComment
\SetFuncSty{textsc}

\SetKw{Skip}{skip}
\SetKw{Break}{break}
\SetKw{Continue}{continue}

\SetKwFunction{HasCandidates}{HasCandidates}
\SetKwFunction{ExtractCandidate}{ExtractCandidate}
\SetKwFunction{IsConforming}{IsConforming}
\SetKwFunction{ConstructPrompt}{ConstructPrompt}
\SetKwFunction{ExtractDafnyCode}{ExtractDafnyCode}
\SetKwFunction{PromptLLMForPatches}{PromptLLMForPatches}
\SetKwFunction{PrioritizeIntent}{PrioritizeIntent}

\KwIn{Hard intent $\phi_{H}$, Soft intent $\phi_{S}$, Failing trace $t_\mathit{fail}$, Number of patches $k$}
\KwOut{Patches \textit{$Ps$}}

$\phi_{S}' \gets $ update $\phi_{S}$ with priorities for each fact;

$\phi_{S}'' \gets $ retrieve from $\phi_{S}'$ the facts with highest priority;

$\program_a,\spec_a \gets$ translate $\phi_{H}$, $\phi_{S}$, $t_\mathit{fail}$ back to Dafny;

$\program'',\spec'' \gets$ translate $\phi_{S}''$ back to Dafny;

$Q' \gets$ construct a prompt using $\program_{a}$, $\spec_{a}$, $\program''$, $\spec''$;

$Ps \gets$ ask LLM for $k$ patches using prompt $Q'$;
\end{algorithm}
\end{minipage}
\end{wrapfigure}
 $\algo{SynthesizePatch}$. 
Selecting to patch one or more facts from the {soft intent} can lead to a successful repair for the whole program. 
To prioritize the repair of the facts that would produce the least intrusive fix for program-specification conformance, we order all the facts lexicographically over a triple of the following three criteria (lines 1):
\begin{enumerate*}
    \item number of non-conforming facts from the hard intent;
    \item number of non-conforming facts from the soft intent;
    \item strength of formulae.
\end{enumerate*}
We next select to fix the facts with the highest priority (line 2), that is, the facts removing ``most'' non-conformance.
For patch synthesis, we leverage the power of a Large Language Model synthesizer. This enables \toolName to tackle multi-hunk repairs for a single failure and to repair both the program $I_{fail}$ and specification $(\pre,\spec_\mathit{fail})$ simultaneously, leading to a real \emph{co-evolution} towards conformance. 
Before prompting the model, we relate the intent and trace back to the original Dafny program $\program_a$ and specification $\spec_a$ to present the problem in a more typical setup for an LLM (line 3). 
In the process of translating back to Dafny, \toolName also annotates the statements and specifications corresponding to the hard intent with a {\mintinline{C}{:trusted}} attribute, instructing the model not to modify them.
We also translate back the prioritized soft intent $\phi_{S}''$ as a hint for the model to focus on repairing the corresponding $\program''$ and $\spec''$ fragments (line 4). 
We construct a prompt using the original but now annotated program and specification, $\program_a$ and $\spec_a$, respectively--along with the highest-priority facts and use it to synthesize $k$ patches (lines 5 and 6).

{\bf Co-evolution.}
\autoref{alg:coEvolution}
describes our conformance strategy.
It iteratively identifies and fixes 
\RestyleAlgo{ruled}
\begin{wrapfigure}[16]{r}{0.58\linewidth}
\vspace{-1.5em}
\setlength{\algomargin}{8pt}
\begin{minipage}{0.58\textwidth}
\begin{algorithm}[H]
\caption{\algo{CoEvolution}}
\label{alg:coEvolution}
\small
\DontPrintSemicolon
\LinesNumbered
\SetNoFillComment
\SetFuncSty{textsc}
\SetKw{Skip}{skip}
\SetKw{Break}{break}
\SetKw{Continue}{continue}
\SetKwFunction{HasCandidates}{HasCandidates}
\SetKwFunction{ExtractCandidate}{ExtractCandidate}
\SetKwFunction{ApplyPatch}{ApplyPatch}
\SetKwFunction{IsConforming}{IsConforming}
\SetKwFunction{GetFailingTrace}{GetFailingTrace}
\SetKwFunction{ExtractHardSoftIntent}{ExtractHSIntent}
\SetKwFunction{SynthesizeCandidate}{SynthesizeCandidate}
\SetKwFunction{SynthesizePatches}{SynthesizePatches}
\SetKwFunction{HasProgress}{HasProgress}
\SetKwFunction{CorrectSpec}{CorrectSpec}
\KwIn{Program $\program$, Specification $\spec$}
\KwOut{Verifiable solutions \textit{$V_p$}}

$C \gets \{  (\program,\spec) \};$

\While{ $C \neq \{\}$ }
{
    $ (\program,\spec) \gets $ extract a candidate from $C$
    
    \uIf{$\conformsPS{\program}{\spec}$}{
        $V_p \gets V_p \cup \{ (\program,\spec) \};$

        \Continue;
    }
    $ t_{\mathit{fail}} \gets$ get failing trace for $(\program,\spec);$
    
    $\phi_H , \phi_S \gets \ExtractHardSoftIntent(\program, \spec);$
    
    $P \gets \SynthesizePatches(\phi_H, \phi_S,t_{\mathit{fail}},k);$

    \For{ ${(\program''',\spec''')} \in {P}$ }
    {
    $(\program_r,\spec_r) \gets$  apply patch $(\program''',\spec''')$ on $(\program,\spec)$;

    $C \gets$ add $(\program_r,\spec_r)$ to $C$ if it does not exhibit $t_\mathit{fail}$ ;
    }
}
\end{algorithm}
\end{minipage}
\end{wrapfigure}
all non-conforming partitions, each represented by a failing trace, until complete conformance is 
 achieved, resulting in conforming program specification pairs collected in $V_p$.
Starting with $(\program,\spec)$ (line 1), the original pair of program and specification, the set $C$ iteratively collects different versions of refined pairs of program and specification until there are no more non-conforming pairs to repair (line 2) or until the upper bound of allowed iterations is reached (not shown in the algorithm). 
If the candidate pair (line 3) is conforming, it is added to the set of verifiable pairs, $V_p$ (line 4-5).
Otherwise, that is, if $\notconforms{\program}{\spec}$, 
\toolName picks one failing trace $t_\mathit{fail}$ to be repaired (line 7).
When selecting a failing trace, we prioritize the returned traces in a reverse trace-depth ordering to focus on the non-conformance closer to the start of the failing program. 
Next, to prepare for the synthesis of a patch for $t_\mathit{fail}$,
 \toolName extracts the hard and soft intent from $\program$ and $\spec$, respectively, using \autoref{alg:intent-extraction} and passes this information along with an integer $k$ representing the number of patches to be synthesized to the synthesizer described by \autoref{alg:synthesis}. 
All the patches $(\program''',\spec''')$ returned by the synthesizer are applied to the original pair, resulting in a refined program and specification pair $(\program_r,\spec_r)$ (line 11).
If the refined pair no longer exhibits the failing trace $t_\mathit{fail}$ it is added to the set of 
refined pairs of program and specification (line 12) to be further refined in a subsequent iteration of the refinement loop (line 2).

\subsection{P2: Aligning Specification with Programmer's Intent Through Tests}
\label{subsec: AlignIntent}

\RestyleAlgo{ruled}
\begin{wrapfigure}[17]{r}{0.61\linewidth}
\vspace{-1em}
\setlength{\algomargin}{8pt}
\begin{minipage}{0.61\textwidth}
\begin{algorithm}[H]
\caption{\algo{AutomatedAssurance}}
\label{alg:assuranceCycle}
\small
\DontPrintSemicolon
\LinesNumbered
\SetNoFillComment
\SetFuncSty{textsc}
\SetKw{Skip}{skip}
\SetKw{Break}{break}
\SetKw{Continue}{continue}
\SetKwFunction{Coevolution}{CoEvolution}
\SetKwFunction{AutomatedAssurance}{AutomatedAssurance}
\KwIn{Program $\program_{0}$, Specification $\spec_{0}$, Test suite $\test$}
\KwOut{Conforming triples {$V_t$}}

$V_t \gets \{\}$

$V_p \gets \Coevolution(\program_{0},\spec_{0});$

$\spec_{\test} \gets $ translate a test suite $\test$ to specification format;

\For{ ${(\program,\spec)} \in {V_p}$ }{
    $\program_{\spec} \gets $ translate $\spec$ to a Dafny code format;
    
    $V_p^r \gets \Coevolution(\program_{\spec},\spec_{\test});$
    
    \For{ ${(\program^r_{\spec},\spec^r_{\test})} \in {V^r_p}$ }{

        $\spec_r \gets$ translate $\program^r_{\spec}$ to specification format 

        $\test_r \gets$ translate $\spec^r_{\test}$ to test format 
        
        \uIf{$\conformsPS{\program}{\spec_r}$}{
                $V_t \gets V_t \cup \{(\program,\spec_r,\test_r) \}$
        }
        
        \uElse{
        $V_t \gets V_t~ \cup $ \AutomatedAssurance$(\program,\spec_r,\test_r)$ 
        }
    }
}
\end{algorithm}
\end{minipage}
\end{wrapfigure}
\autoref{alg:coEvolution} offers us a solution for 
problem {\bf P1}. In other words, we have an updated collection $ V_{p} = \{ (\program, \spec) : (\conformsPS{\program}{\spec})  \}$ of repaired 
programs and specifications. Although these pairs are conforming, they might not align with the programmer's intent. 
To achieve alignment with the programmer's intent, we support  ``interaction'' through test cases. 
A user is offered a test that they can modify to express their intent. 
These tests can originate from a trusted or untrusted oracle, such as a user or an LLM.
\autoref{alg:assuranceCycle}, $\mathtt{AutomatedAssurance}$ presents a methodology to achieve alignment between a conforming pair and a test.
We prioritize the alignment specifications with tests over the alignment of programs with tests, as the latter aligns with more common verification use cases.

Once the program $\program_{0}$ and specification $\spec_{0}$ co-evolution campaign has ended with a set of verified pairs $V_{p}$ (line 2), we prepare for another campaign that takes into consideration tests. 
The first step in the preparation is to translate $T$ into a specification $\spec_{\test}$ (line 3) as described in \autoref{sec:background}. 
While iterating over all pairs ($\program,\spec$) in $V_{p}$ (line 4), we translate the specification $\spec$ into a program $\program_{\spec}$ (line 5) before running a co-evolution campaign over the specification $\spec$ and test $T$ represented by the pair $(\program_{\spec}, \spec_{T})$ (line 6).
If a candidate ($\program_{\spec}^{r},\spec_{\test}^{r}$) resulting from the co-evolution campaign conforms with program $\program$,  we store the fully-conforming triple $(\program,\spec_{r}, \test_{r})$ into the set $V_{T}$ (lines 10 and 11).
Before doing this check, $\program_{\spec}^{r}$ is translated to a specification $\spec_{r}$, and $\spec_{\test}^{r}$ to a test $\test_{r}$ (lines 8 and 9).
If it does not conform, another cycle of the algorithm is initiated with the triple $(\program,\spec_{r}, \test_{r})$ with refined specification and test (line 13). We discuss termination conditions in \autoref{sec:evalution}.

\section{Evaluation}\label{sec:evalution}
The goal of \toolName is to construct valid artifacts that are verifiable programs while disambiguating the programmer's intent. We examine the following research questions.
\begin{itemize}
\item[\bf RQ1:] How effective is program-proof co-evolution in constructing conforming artifacts? 
\item[\bf RQ2:] What is the quality of the repaired formal postconditions? 
\item[\bf RQ3:] Does the refined, informal natural language intent reduce the ambiguity in the user intent? 
\end{itemize}

\subsection{Setup}\label{sec:setup}
We implement our approach in a tool named \toolName. 
The main driver is about 1000 lines of Python code that implements the assurance cycle and prompts the model. 
We also extend Boogie to generate the hard and soft intents and extract the underlying failing assertion.
The invocation of Boogie is done through Dafny. 
The extension of these tools comes down to about 500 lines of C\#.

{\bf Prompt.}
\toolName uses the following system prompt for interacting with the LLM-based synthesizer (node \textbf{5} in \autoref{fig:coevolution}):

\begin{tcolorbox}[size=small,boxrule=0pt,title=System Prompt for \toolName,width=\textwidth,fonttitle=\footnotesize,]
\footnotesize
You are an expert Dafny programmer.
You know Dafny's verification process very well.
You know how to explicitly define Dafny invariants in loops.
You know how to write preconditions and postconditions for methods.
You know how to write Dafny lemmas and functions.

\textbf{Refresher}: <Dafny basics about its semantics and verification process>

\textbf{Goal}: Your task will be to repair a given Dafny program that has one or more verification errors. You have to repair one of the verification errors such that the program no longer has this error. Use your knowledge to do that while following the guidelines given to you below under the Guidelines header.

\textbf{Input prompt format}: As input, you will be given seven parts in the following format: ...

\textbf{Output prompt format}: As output, write a patch for the verification error. Return the patch in the format below ...

\textbf{Guidelines}: <Semantics of hard and soft intent annotations and additional goals for generating a repair>
\end{tcolorbox}

We use the system prompt above to define the problem setup for the model and express the full range of capabilities in Dafny. 
The Refresher section presents rules specific to Dafny, such as information about ghost data and lemmas. 
The Input and Output format sections describe the exact format in which we will prompt the model and the output of the patches it must generate.
The Guidelines section presents repair specific rules, such as the requirement that the synthesizer must maintain the annotated hard intent and recommendations for what part of the soft intent to repair.

For the naive setup, we use the following system prompt:

\begin{tcolorbox}[size=small,boxrule=0pt,title=System Prompt for naive repair,width=\textwidth,fonttitle=\footnotesize,]
\footnotesize
You are a programming assistant in Dafny, with program repair expertise ...
\end{tcolorbox}

{\bf Dataset.} We use the recently proposed dataset $\mathtt{MBPP\!-\!DFY}$ by Misu et al.~\cite{misu2024towards}. 
The dataset consists of a subset of 178 problems from the MBPP dataset~\cite{mbpp}, which are used to assess the capability of the Large Language Models to generate verifiable Dafny code. The dataset contains
\begin{wrapfigure}[6]{r}{0.58\linewidth}
\vspace{-0.8em}
\begin{minipage}{0.58\textwidth}
\includesvg[width=1\textwidth]{figures/problemDistribution}
\end{minipage}
\end{wrapfigure}
1,054 LLM-generated Dafny programs corresponding to 178 problem statements. Of these, 620 programs do not compile and are therefore excluded from our evaluation.
From the remaining 434 programs that do compile, we extract 60 subjects whose program and specification do not conform and use them to evaluate \toolName's \emph{program-specification conformance repair capability}. 
We call this collection of 60 subjects {\bf datasetNV}.
Furthermore, 48 subjects of the 374 verifying programs have syntactically valid tests that do not conform with the specifications and program, and 326 have syntactically invalid tests.
From the 326 subjects, we semi-automatically repaired the syntactic problems and extracted 205 non-conforming exemplars. 
We call this collection of 48 + 205 = 253 subjects that have programs and specifications conforming but not tests, as  {\bf datasetV}. 
The remaining 121 subjects verify, hence we exclude them from our evaluation.
We use {datasetV} to evaluate \toolName's \emph{program-specification-test conformance repair capability}.

{\bf Configuration.} To evaluate \toolName and the baselines in different setups, we use two models -- OpenAI's \gptfouro (gpt-4o-2024-05-13) and Anthropic's
\sonnet (claude-3-5-sonnet-20240620).
We selected these models as they are state-of-the-art for code generation at the time of writing. 
To explore the models' capabilities for repairing Dafny code, a language less represented in training sets, we use three different temperatures, $T \in \{0.3,0.7,1.0\}$. 
To ensure termination, we have set up multiple bounds for \toolName:
\begin{enumerate*} 
\item time budget of 20 minutes per subject;
\item maximum of 5 co-evolution campaigns;
\item stop if a verifying candidate is found within the above bounds.
\end{enumerate*}
We give the baseline five attempts to repair a non-verifiable program. 

\subsection{RQ1: Reaching Conformance}
In this research question, we aim to evaluate the capability of \toolName to repair Dafny programs. We examine program-specification conformance using {\bf datasetNV} and program-specification-test conformance repair using {\bf datasetV}.
\paragraph{Experimental Setup}

We compare \toolName with two baselines. The first one uses a simple prompt to ask the LLM to repair the conformance issue given a collection of verification failures as input. We call this the ``\naiverepair''. The second tool enhances the \naiverepair~ and uses the same system prompt as the one \toolName uses. We call this the ``\chainofthought''. 

\paragraph{Results}

\begin{table}[t]
\small 
    \centering
    \resizebox{0.85\textwidth}{!}{  
    \begin{tabular}{lccccccl}
    \toprule 
    &  & \multicolumn{2}{c}{\naiverepair} & \multicolumn{2}{c}{\chainofthought} & \multicolumn{2}{c}{\toolName}\\
    \cmidrule(lr){3-4}\cmidrule(lr){5-6}\cmidrule(lr){7-8}
               Model & Temp. $T$ & Aligned & Avg. Time  & Aligned & Avg. Time & Aligned & Avg. Time\\
   \midrule
    \gptfouro & 0.3 & 11 & 43s & 15  & 55s & 23 & 111s  \\ 
              & 0.7 & 13 & 46s & 15  & 55s & 23 & 81s \\  
              & 1.0 & 13 & 39s & 13  & 58s & 22 & 160s  \\
    \midrule
    \sonnet  & 0.3 & 12 & 45s & 18  & 74s  & 27 & 249s \\
             & 0.7 & 12 & 46s & 21  & 73s  & 31 & 236s \\
             & 1.0 & 14 & 49s & 24  & 87s  & 30 & 252s \\ 
    \bottomrule
    Average & & 12.8 (21.3\%) & 45s & 17.3 (28.9\%)  & 67s  & 22.2 (37\%) & 182s \\ 
    \end{tabular}
 }   
\vspace{1em}
\caption{Reaching conformance between program and specification on {\bf datasetNV} (60 subjects) }
\vspace{-1em}
\label{tab:rq1-1}
\end{table}

\autoref{tab:rq1-1} summarizes our results for evaluating conformance between a program and a specification. 
Across models, we observe that \toolName aligns more programs with specifications than both baselines. 
For example, using \sonnet with $T=1.0$, \toolName aligns 30 programs which is twice the amount of programs aligned by the \naiverepair~ which repairs only 14, and 25\% more than the \chainofthought~ which repairs 24. 
Though the average time is significantly higher for \toolName, we attribute this to the larger amount of model calls resulting from the incremental repair process and the additional calls to Dafny required for extracting intent. 
Furthermore, due to Anthropic's APIs not providing multiple responses from a prompt, the average tool time is higher when using \sonnet\!\!\!.
Lastly, a manual inspection of all outputs revealed that the difference in effectiveness between models comes from the model's capability to follow the instructions in the prompts.
In particular, \gptfouro dismisses some guidelines more often than \sonnet does. 
\result{RQ1.1: \toolName  aligns about 73\% and 28\% more programs with specifications than a \naiverepair~ baseline and a \chainofthought~ repair, respectively. }

\begin{table}[t]
\small
    \centering
    \resizebox{0.85\textwidth}{!}{  
    \begin{tabular}{lccccccl}
    \toprule 
    &  & \multicolumn{2}{c}{\naiverepair} & \multicolumn{2}{c}{\chainofthought} & \multicolumn{2}{c}{\toolName}\\
    \cmidrule(lr){3-4}\cmidrule(lr){5-6}\cmidrule(lr){7-8}
               Model & Temp. $T$ & Aligned & Avg. Time  & Aligned & Avg. Time & Aligned & Avg. Time\\
   \midrule
    \gptfouro & 0.3 & 66  & 45s & 66  & 58s & 69 & 94s  \\
              & 0.7 & 69  & 48s & 66  & 64s & 71 & 70s \\
              & 1.0 & 77  & 40s & 77  & 58s & 75 & 67s  \\
    \midrule
    \sonnet & 0.3 & 67 & 48s & 69 & 60s & 73 & 143s \\
            & 0.7 & 68 & 49s & 68 & 62s & 79 & 140s \\
            & 1.0 & 69 & 46s & 78 & 63s & 88 & 195s \\
    \bottomrule
    Average &  &  69.3 (27.3\%) & 46s & 70.6 (27.9\%)  & 61s & 75.83 (30\%)  & 118s  \\
    \end{tabular} }
    \vspace{1em}
\caption{Reaching conformance between program, specification and tests on {\bf datasetV} (253 subjects) } \vspace{-2em}
\label{tab:rq1-2}
\end{table}

\autoref{tab:rq1-2} summarizes our results for evaluating conformance between a program, a specification, and a test. 
Across models, we observe that \toolName aligns more programs with specifications and tests than both baselines.
Compared to the results of program-specification alignment, we see that the effect of the \chainofthought~ system prompt is less significant, where \toolName can increase the alignment percentage by only two percentage points. 
A manual examination of the failures revealed three main reasons for the inability of \toolName and the baselines to repair more subjects. 
The main reason is the models' difficulty of inferring or strengthening specifications, even for the trivial programs in the dataset. 
Another weakness is the models' inability to weaken the preconditions when presenting a test that uses input data that does not satisfy the constraints. 
Furthermore, even though the \chainofthought~ repair tool and \toolName explicitly define the concept of triggers for specifications with quantifiers, the models do not generate them often enough to support the verification of aligned subjects.

After manually examining the results of \sonnet with temperature $T=1.0$ we observe that for the problems no other instance solved, \toolName has successfully aligned the artifacts by constructing lemmas or using intermediate assertions to invoke quantifier triggers.
We attribute this capability to the high temperatures, allowing the model to try more uncommon patches.
We believe that with an increased number of candidates sampled, the number of solutions can increase.

\result{RQ1.2: \toolName on average performs similarly to a \naiverepair ~ baseline  and a \chainofthought~ repair for aligning program-specifications-tests. }

\subsection{RQ2: Behavior Coverage }
In this research question, we aim to evaluate the capability of \toolName to repair Dafny programs to reach a higher quality postcondition. We use the completeness evaluation metric proposed by Lahiri~\cite{lahiri2024evaluatingllmdrivenuserintentformalization}. The completeness score proposed by Lahiri quantifies the strength of a postcondition $\varphi$ based on a set of tests $T$, representing consistent input-output pairs. The completeness score is defined as the fraction of output mutations from $T$ that are inconsistent with $\varphi$. This metric is inspired by the kill-set concept in mutation testing and intuitively reflects how well the postcondition differentiates the conforming code from the incorrect alternatives, the higher the score the better.

\paragraph{Experimental setup}
Using {\bf datasetNV}, we examine the postcondition completeness of the program at three stages - before program-specification conformance, after program-specification conformance, and after program-specification-test conformance. Using {\bf datasetV}, we examine the postcondition completeness score from a generated subject that is aligned and the score from the same subject after being aligned with tests. 
We use a mutation set of size 20 in this experiment.
\paragraph{Results}

\autoref{tab:rq2-1} summarizes our results for evaluating the capability of \toolName
to increase
 the quality of the postcondition while reaching conformance.
Column ``Initial'' denotes the initial value of the dataset before executing \toolName. Columns ``PS'' and ``PST'' represent program-specification conformance and program-specification-test conformance, respectively.
We observe 
\begin{wraptable}[10]{r}{80mm}
    \centering
    \resizebox{0.55\textwidth}{!}{  
    \begin{tabular}{lcccccl}
    \toprule 
    &  & \multicolumn{2}{c}{{\bf datasetNV}} & \multicolumn{2}{c}{{\bf datasetV}}\\
    \cmidrule(lr){3-4}\cmidrule(lr){5-6}
               Model & Temperature $T$ & Initial (None) & PS & Initial(PS) & PST \\
    \midrule           
    \gptfouro & 0.3 & 0.38 & 0.49 &  0.28  & 0.89 \\
              & 0.7 & 0.42 & 0.41 &  0.31  & 0.88 \\
              & 1.0 & 0.51 & 0.56 &  0.29  & 0.89 \\
    \midrule
    \sonnet   & 0.3 & 0.53 & 0.45 &  0.23  & 0.84 \\
              & 0.7 & 0.50 & 0.61 &  0.22  & 0.86 \\
              & 1.0 & 0.53 & 0.54 &  0.30  & 0.89 \\
    \bottomrule
    Average & & 0.48 & 0.51 & 0.27 & 0.88 \\
    \end{tabular}}
    \vspace{0.7em}
\caption{Evaluating completeness of the postconditions.}
\label{tab:rq2-1}
\end{wraptable}
that for the program-specification conformance, the completeness of the postconditions stays the same on average or increases by up to a maximum of 10\%. 
This shows that \toolName maintains the quality of the provided specifications. 
With the inclusion of tests, the completeness of the postconditions increases significantly since tests ``puts a requirement'' on the specification. 
This exposes the problem of weak specifications, allowing for \toolName to evolve the specifications. We see this with the average increase of 61 percentage points across all setups.

\result{ RQ2: \toolName does not decrease average completeness when conforming a program with specification and increases completeness when all three artifacts are provided by 61pp.}

\subsection{RQ3: Summary ambiguity }


As the intent crystallizes into one that all artifacts (and possibly the user) agree on, we 
see the opportunity of translating it back into natural language for the purpose of using it as program documentation as software evolves or gets rewritten.
Different from the original prompt, this should be a clearer and unambiguous description of the desired computation. We, therefore, assess the ambiguity of this description.

\paragraph{Experimental setup}
\begin{wrapfigure}[15]{r}{0.5\textwidth}
\vspace{-2.5em}
\hspace{-1.7em}
  \includegraphics[width=0.6\textwidth]{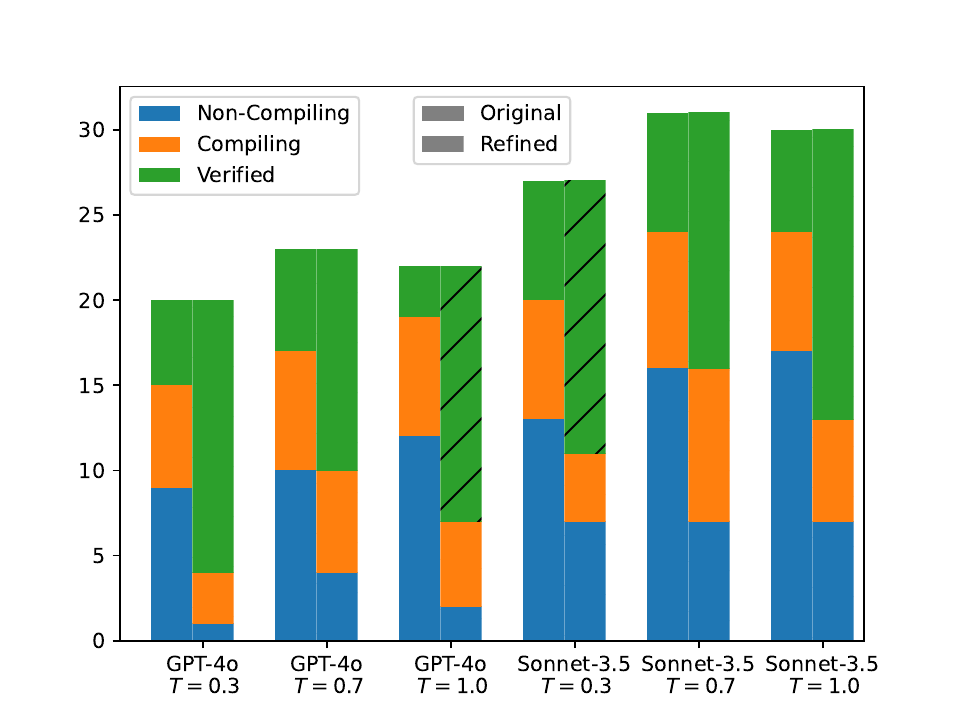}
    \vspace*{-8mm}
    \caption{Assessing ambiguity of intent.}
  \label{tab:rq3}
\end{wrapfigure}
We use the aligned artifacts created by \toolName for {\bf datasetNV}, 
specific to the selected LLMs and their varying temperature settings, to extract the aligned summary 
\noindent and assess its ambiguity.  
First, we convert the identified formal intent into a natural language description. We then use this description as a prompt for the LLM to generate verifiable code. 
If the generated code successfully passes verification, it indicates that the summary is clear and unambiguous, which reduces nondeterminism in code generation and reflects a strong alignment with the desired intent. 

For the translation from formal intent to natural language, we annotate each verifiable code produced by \toolName with {\mintinline{C}{:trusted}} to indicate hard intent. We then prompt the model to summarize the annotated code, instructing it to focus on the computation marked by the annotations. 
This approach guides the summarisation to focus on the \emph{what} instead of the \emph{how}. 

\paragraph{Result}
\autoref{tab:rq3} provides a summary of the results from our experiment across different LLMs and temperatures. The ``Original'' and ``Refined'' bars reflect the outcomes of using the original problem statement and the aligned program summary, respectively, as prompts for producing verified code.
We observe a substantial increase in the number of programs that successfully compile (the Compiling and Verified sub-bars) and those that verify correctly (the Verified sub-bar) when using the aligned summary. 
This enhancement can be attributed to the aligned summaries offering essential information on preconditions, high-level algorithms, and invariants, thereby enabling the language models to concentrate on accurately expressing the program's behavior.

\result{ RQ3: The summaries produced from an aligned program disambiguate the intent and allow generating the same program when provided to a different model. }

\section{Discussion}
\label{sec:discussion}
\ignore{We have presented an implementation based on the Dafny language and Boogie framework, but the approach does not bind us to any specific programming language. The only requirement is that the language be expressable in a system that can verify and extract a counterexample trace. 

Our current work does not focus on the problem of class invariants and their repair. In that setting, we can observe that an invariant must simultaneously be satisfied by many methods if selected for repair. Thus, we can define the cost of modifying the invariant and construct a complex synthesis specification to construct a correct invariant.

Verification has limitations and synthesis has limitations. Using Dynamic Execution of tests can increase the capabilities of conformance checking between program/spec and test.

As we are using large language models for synthesis, there are risks of the model continuously applying to set of patches that reverse one another. To ensure against this we also define the patched code as hard intent. \mm{possibly should be in section 4?} 

Enhancing the partitioning through data dependency analysis?}

{\bf Limitations.}
Our approach currently tackles one failing trace at a time. We leave multipath repair by merging failing traces as future work, where one pass of the co-evolution process fixes the conformance problem entirely. 
We currently assume that any lemma required by the verification is already provided and is proven correct. Future work could look into leveraging LLMs to infer required lemmas or repair their body when not proven correct.
The current limitations stem from our reliance on the Dafny ecosystem. While we are hopeful about discovering user intent in other programming languages through a co-evolution of artifacts and refinement of hard and soft intents, we recognize that new programming environments may present their own set of challenges.

{\bf Threats to Validity.} 
\emph{Internal.} An LLM-based synthesiser is not complete, hence it may not always find a solution even when one exists. 
Furthermore, LLMs are prone to hallucinations, which can result in patches that do not fully adhere to the constraints given to the synthesizer, such as maintaining the hard intent. These solutions are filtered out through an additional validation step where Boogie verifies that the hard intent still holds for the patched Dafny program. A further improvement would be to translate the hard intent and other constraints into unit tests, further filtering out solutions that violate the synthesizer's constraints. 
To handle quantifiers, Dafny infers what are known as triggers \cite{dafnyfaq}, which serve as additional hints for instantiating quantified formulae. However, these triggers might not always be automatically inferred, leading to situations where a program is correctly patched but still not verifiable. In such cases, manual intervention is required from an oracle to specify the necessary triggers.

\emph{External.} One external threat is the reliance on the combination of the $\mathtt{MBPP\!-\!DFY}$ dataset and our selected models for the evaluation. At the time of writing this submission, this combination represents the state of the art in demonstrating the capability of LLMs to synthesize Dafny programs in the absence of program repair interventions. However, as the models become more powerful, the effectiveness gap between the baseline for achieving program-specification conformance and our approach may narrow and the risks of data leakage can increase. 
Nevertheless, intent disambiguation is likely to remain a challenge in the foreseeable future due to LLMs' inherent dependence on natural language prompts and the tendency of users to provide insufficiently precise intent descriptions.

\ignore{
\begin{itemize}
    \item Internal. 
    : the synthesis is LLM based (Using a synthesis engine oriented towards the logical formulae.), data leakage
    \item External: dataset (extendability to other Dafny programs, but also to other languages), model, reliance on hard intent (that is, the model can generate almost correct solutions)
\end{itemize}
}

\section{Related Work}
\label{sec:relatedWork}
A significant body of work in automated program repair ~\cite{semfix,directfix,extractfix} focuses on synthesizing patches by gathering constraints mainly derived from the provided unit tests. 
Various methodologies like static analysis ~\cite{Sharon2007}, abstract interpretation ~\cite{Cousot2012}, and dynamic analysis ~\cite{Ernst99} have been used to automatically infer specifications from code.
Recent studies ~\cite{hahn2022formal,pmlr-v202-pei23a,kamath2024finding,endres2024can} have also explored the utilization of large language models (LLMs) for generating formal specifications in different formats.
For instance, Kamath et al. ~\cite{kamath2024finding} demonstrated that LLMs can generate and repair inductive invariants, while Chakraborty et al. ~\cite{chakraborty2023ranking} introduced an approach to rerank the generated invariants to optimize the number of calls to the verifier. Endres et al. ~\cite{endres2024can} proposed a method for creating postconditions from natural language definitions, which aids in identifying code errors for potential repairs. Additionally, Hahn et al. ~\cite{hahn2022formal} examined the generalization capabilities of language models by fine-tuning the T5 model to generate competitive Linear Temporal Logic formulas, First-Order Logic, and regular expressions. Pei et al. ~\cite{pmlr-v202-pei23a} further fine-tuned a language model with invariants mined from the dynamic analysis tool Daikon, enabling the inference of new invariants in a single step or across multiple execution points.
Building on the insights from these studies, we developed a co-evolution strategy that enables the simultaneous repair of code, invariants, preconditions, and postconditions in an integrated manner.

The work by First et al. ~\cite{first2023baldur} presents an approach in which an LLM is fine-tuned to repair proofs in conjunction with another LLM-based tool, Thor ~\cite{jiang2022thor}, for generating correct code in Isabelle/HOL. In contrast, Gopinathan et al. ~\cite{gopinathan2023mostly} modify proofs to align with newly created code, assuming that the program's evolution is limited to non-functional changes. LeanDojo ~\cite{yang2023leandojo} focuses on extracting data from Lean and enabling programmatic interaction with the proof environment, providing fine-grained annotations of premises in proofs that are crucial for premise selection—a key challenge in theorem proving. Ringer's work ~\cite{ringer2021proof} centers on proof repair within an inductive reasoning system, aiming to maintain the proof while evolving system requirements without altering those requirements. In contrast to these approaches, our workflow allows for the simultaneous evolution of code alongside the proofs and specifications.

Similar to our work, Misu et al.~\cite{misu2024towards} choose the Dafny programming language and utilize 178 programming problems from the MBPP dataset to prompt the models GPT-4 and PaLM-2 for generating methods in Dafny.
They demonstrate that employing a Chain of Thought prompt can effectively produce verified and accurate Dafny methods with meaningful specifications. 
In another recent study, Sun et al. ~\cite{sun2023clover} focus on iterating code generation without applying any repairs, using artifacts like comments and existing annotations as feedback to produce correct outputs, emphasizing the importance of internal consistency.
Huang et al. ~\cite{huang2024enhancinglargelanguagemodels} also tackle the challenge of aligning intent from specifications, code, and test cases, aiming to rerank solutions through intra- and inter-consistency. 
In our evaluation, we uniquely use tests as a hard constraint to represent the user's intent, guiding the entire evolution process. 
Furthermore, rather than merely ranking the solutions provided by the language model, we adopt an iterative co-evolution approach to refine and develop all three artifacts—specifications, code, and tests—according to the user's intent.

\section{Outlook}
\label{sec:conclusion}

With the advent of LLMs in automatic programming, the interest in trusted automatic programming via LLMs increases. Unfortunately, it is difficult to give any guarantees about code generated from LLMs, partly also because a detailed specification of the intended behavior is usually not available. In this paper we alleviate this lack of functionality specifications by aligning automatically generated code via LLMs, automatically generated formal specifications (obtained from natural language using LLMs), as well as tests. The conformance between generated programs, generated specifications, and tests - does not provide absolute guarantees but enhances trust. Establishing such conformance also helps us uncover the likely intended program behavior.

In this work, we developed this theme of program-proof co-evolution, with the goal of enhancing trust in automatically generated code. There can be other use-cases of our program-proof co evolution technology, such as automatic documentation generation and specific scenarios of software maintenance where a new software component needed by a larger software project may undergo our program-proof co-evolution methodology. 



\bibliographystyle{ACM-Reference-Format}
\bibliography{main}


\begin{thebibliography}{32}


\ifx \showCODEN    \undefined \def \showCODEN     #1{\unskip}     \fi
\ifx \showDOI      \undefined \def \showDOI       #1{#1}\fi
\ifx \showISBNx    \undefined \def \showISBNx     #1{\unskip}     \fi
\ifx \showISBNxiii \undefined \def \showISBNxiii  #1{\unskip}     \fi
\ifx \showISSN     \undefined \def \showISSN      #1{\unskip}     \fi
\ifx \showLCCN     \undefined \def \showLCCN      #1{\unskip}     \fi
\ifx \shownote     \undefined \def \shownote      #1{#1}          \fi
\ifx \showarticletitle \undefined \def \showarticletitle #1{#1}   \fi
\ifx \showURL      \undefined \def \showURL       {\relax}        \fi
\providecommand\bibfield[2]{#2}
\providecommand\bibinfo[2]{#2}
\providecommand\natexlab[1]{#1}
\providecommand\showeprint[2][]{arXiv:#2}

\bibitem[Austin et~al\mbox{.}(2021)]%
        {mbpp}
\bibfield{author}{\bibinfo{person}{Jacob Austin}, \bibinfo{person}{Augustus Odena}, \bibinfo{person}{Maxwell Nye}, \bibinfo{person}{Maarten Bosma}, \bibinfo{person}{Henryk Michalewski}, \bibinfo{person}{David Dohan}, \bibinfo{person}{Ellen Jiang}, \bibinfo{person}{Carrie Cai}, \bibinfo{person}{Michael Terry}, \bibinfo{person}{Quoc Le}, {and} \bibinfo{person}{Charles Sutton}.} \bibinfo{year}{2021}\natexlab{}.
\newblock \bibinfo{title}{Program Synthesis with Large Language Models}.
\newblock
\newblock
\showeprint[arxiv]{2108.07732}~[cs.PL]
\urldef\tempurl%
\url{https://arxiv.org/abs/2108.07732}
\showURL{%
\tempurl}


\bibitem[Barnett and Leino(2005)]%
        {barnett2005weakest-precondition}
\bibfield{author}{\bibinfo{person}{Mike Barnett} {and} \bibinfo{person}{Rustan Leino}.} \bibinfo{year}{2005}\natexlab{}.
\newblock \showarticletitle{Weakest-Precondition of Unstructured Programs}. In \bibinfo{booktitle}{\emph{PASTE '05: The 6th ACM SIGPLAN-SIGSOFT workshop on Program analysis for software tools and engineering} (\bibinfo{edition}{paste '05: the 6th acm sigplan-sigsoft workshop on program analysis for software tools and engineering} ed.)}. \bibinfo{publisher}{ACM Press}, \bibinfo{pages}{82--87}.
\newblock
\showISBNx{1-59593-239-9}
\urldef\tempurl%
\url{https://www.microsoft.com/en-us/research/publication/weakest-precondition-of-unstructured-programs/}
\showURL{%
\tempurl}


\bibitem[Chakraborty et~al\mbox{.}(2023)]%
        {chakraborty2023ranking}
\bibfield{author}{\bibinfo{person}{Saikat Chakraborty}, \bibinfo{person}{Shuvendu Lahiri}, \bibinfo{person}{Sarah Fakhoury}, \bibinfo{person}{Madan Musuvathi}, \bibinfo{person}{Akash Lal}, \bibinfo{person}{Aseem Rastogi}, \bibinfo{person}{Nikhil Swamy}, {and} \bibinfo{person}{Rahul Sharma}.} \bibinfo{year}{2023}\natexlab{}.
\newblock \showarticletitle{Ranking LLM-Generated Loop Invariants for Program Verification}. In \bibinfo{booktitle}{\emph{2023 Empirical Methods in Natural Language Processing}}. \bibinfo{address}{Singapore}.
\newblock
\newblock
\shownote{EMNLP-Findings 2023}.


\bibitem[Cousot et~al\mbox{.}(2012)]%
        {Cousot2012}
\bibfield{author}{\bibinfo{person}{Patrick~M. Cousot}, \bibinfo{person}{Radhia Cousot}, \bibinfo{person}{Francesco Logozzo}, {and} \bibinfo{person}{Michael Barnett}.} \bibinfo{year}{2012}\natexlab{}.
\newblock \showarticletitle{An abstract interpretation framework for refactoring with application to extract methods with contracts}. In \bibinfo{booktitle}{\emph{Proceedings of the ACM International Conference on Object Oriented Programming Systems Languages and Applications}} (Tucson, Arizona, USA) \emph{(\bibinfo{series}{OOPSLA '12})}. \bibinfo{publisher}{Association for Computing Machinery}, \bibinfo{address}{New York, NY, USA}, \bibinfo{pages}{213–232}.
\newblock
\showISBNx{9781450315616}
\urldef\tempurl%
\url{https://doi.org/10.1145/2384616.2384633}
\showDOI{\tempurl}


\bibitem[Endres et~al\mbox{.}(2024)]%
        {endres2024can}
\bibfield{author}{\bibinfo{person}{Madeline Endres}, \bibinfo{person}{Sarah Fakhoury}, \bibinfo{person}{Saikat Chakraborty}, {and} \bibinfo{person}{Shuvendu Lahiri}.} \bibinfo{year}{2024}\natexlab{}.
\newblock \showarticletitle{Can Large Language Models Transform Natural Language Intent into Formal Method Postconditions?}. In \bibinfo{booktitle}{\emph{The ACM International Conference on the Foundations of Software Engineering (FSE)}}. \bibinfo{publisher}{ACM}, \bibinfo{address}{Porto de Galinhas, Brazil, Brazil}.
\newblock
\urldef\tempurl%
\url{https://www.microsoft.com/en-us/research/publication/formalizing-natural-language-intent-into-program-specifications-via-large-language-models/}
\showURL{%
\tempurl}
\newblock
\shownote{https://2024.esec-fse.org/details/fse-2024-research-papers/51/Can-Large-Language-Models-Transform-Natural-Language-Intent-into-Formal-Method-Postco}.


\bibitem[Ernst et~al\mbox{.}(1999)]%
        {Ernst99}
\bibfield{author}{\bibinfo{person}{Michael~D. Ernst}, \bibinfo{person}{Jake Cockrell}, \bibinfo{person}{William~G. Griswold}, {and} \bibinfo{person}{David Notkin}.} \bibinfo{year}{1999}\natexlab{}.
\newblock \showarticletitle{Dynamically discovering likely program invariants to support program evolution}. In \bibinfo{booktitle}{\emph{Proceedings of the 21st International Conference on Software Engineering}} (Los Angeles, California, USA) \emph{(\bibinfo{series}{ICSE '99})}. \bibinfo{publisher}{Association for Computing Machinery}, \bibinfo{address}{New York, NY, USA}, \bibinfo{pages}{213–224}.
\newblock
\showISBNx{1581130740}
\urldef\tempurl%
\url{https://doi.org/10.1145/302405.302467}
\showDOI{\tempurl}


\bibitem[et~al.(2024)]%
        {dafnylang}
\bibfield{author}{\bibinfo{person}{Rustan~Leino et al.}} \bibinfo{year}{2024}\natexlab{}.
\newblock \bibinfo{title}{The Dafny Programming and Verification Language}.
\newblock
\newblock
\urldef\tempurl%
\url{https://dafny.org/}
\showURL{%
\tempurl}


\bibitem[Fan et~al\mbox{.}(2023)]%
        {FanICSEFoSE2023}
\bibfield{author}{\bibinfo{person}{A. Fan}, \bibinfo{person}{B. Gokkaya}, \bibinfo{person}{M. Harman}, \bibinfo{person}{M. Lyubarskiy}, \bibinfo{person}{S. Sengupta}, \bibinfo{person}{S. Yoo}, {and} \bibinfo{person}{J.~M. Zhang}.} \bibinfo{year}{2023}\natexlab{}.
\newblock \showarticletitle{Large Language Models for Software Engineering: Survey and Open Problems}. In \bibinfo{booktitle}{\emph{2023 IEEE/ACM International Conference on Software Engineering: Future of Software Engineering (ICSE-FoSE)}}. \bibinfo{publisher}{IEEE Computer Society}, \bibinfo{address}{Los Alamitos, CA, USA}, \bibinfo{pages}{31--53}.
\newblock
\urldef\tempurl%
\url{https://doi.org/10.1109/ICSE-FoSE59343.2023.00008}
\showDOI{\tempurl}


\bibitem[First et~al\mbox{.}(2023)]%
        {first2023baldur}
\bibfield{author}{\bibinfo{person}{Emily First}, \bibinfo{person}{Markus~N. Rabe}, \bibinfo{person}{Talia Ringer}, {and} \bibinfo{person}{Yuriy Brun}.} \bibinfo{year}{2023}\natexlab{}.
\newblock \bibinfo{title}{Baldur: Whole-Proof Generation and Repair with Large Language Models}.
\newblock
\newblock
\showeprint[arxiv]{2303.04910}~[cs.LG]


\bibitem[Gao et~al\mbox{.}(2021)]%
        {extractfix}
\bibfield{author}{\bibinfo{person}{Xiang Gao}, \bibinfo{person}{Bo Wang}, \bibinfo{person}{Gregory~J. Duck}, \bibinfo{person}{Ruyi Ji}, \bibinfo{person}{Yingfei Xiong}, {and} \bibinfo{person}{Abhik Roychoudhury}.} \bibinfo{year}{2021}\natexlab{}.
\newblock \showarticletitle{Beyond Tests: Program Vulnerability Repair via Crash Constraint Extraction}.
\newblock \bibinfo{journal}{\emph{ACM Trans. Softw. Eng. Methodol.}} \bibinfo{volume}{30}, \bibinfo{number}{2}, Article \bibinfo{articleno}{14} (\bibinfo{date}{feb} \bibinfo{year}{2021}), \bibinfo{numpages}{27}~pages.
\newblock
\showISSN{1049-331X}
\urldef\tempurl%
\url{https://doi.org/10.1145/3418461}
\showDOI{\tempurl}


\bibitem[{GitHub}(2021)]%
        {githubCopilot}
\bibfield{author}{\bibinfo{person}{{GitHub}}.} \bibinfo{year}{2021}\natexlab{}.
\newblock \bibinfo{title}{{GitHub Copilot}}.
\newblock \bibinfo{howpublished}{\url{https://copilot.github.com/}}.
\newblock


\bibitem[Gopinathan et~al\mbox{.}(2023)]%
        {gopinathan2023mostly}
\bibfield{author}{\bibinfo{person}{Kiran Gopinathan}, \bibinfo{person}{Mayank Keoliya}, {and} \bibinfo{person}{Ilya Sergey}.} \bibinfo{year}{2023}\natexlab{}.
\newblock \showarticletitle{Mostly Automated Proof Repair for Verified Libraries}.
\newblock \bibinfo{journal}{\emph{Proceedings of the ACM on Programming Languages}} \bibinfo{volume}{7}, \bibinfo{number}{PLDI} (\bibinfo{year}{2023}), \bibinfo{pages}{25--49}.
\newblock


\bibitem[Hahn et~al\mbox{.}(2022)]%
        {hahn2022formal}
\bibfield{author}{\bibinfo{person}{Christopher Hahn}, \bibinfo{person}{Frederik Schmitt}, \bibinfo{person}{Julia~J. Tillman}, \bibinfo{person}{Niklas Metzger}, \bibinfo{person}{Julian Siber}, {and} \bibinfo{person}{Bernd Finkbeiner}.} \bibinfo{year}{2022}\natexlab{}.
\newblock \bibinfo{title}{Formal Specifications from Natural Language}.
\newblock
\newblock
\showeprint[arxiv]{2206.01962}~[cs.SE]


\bibitem[Hoare(1969)]%
        {hoareAxiomatic}
\bibfield{author}{\bibinfo{person}{C.~A.~R. Hoare}.} \bibinfo{year}{1969}\natexlab{}.
\newblock \showarticletitle{An axiomatic basis for computer programming}.
\newblock \bibinfo{journal}{\emph{Commun. ACM}} \bibinfo{volume}{12}, \bibinfo{number}{10} (\bibinfo{date}{oct} \bibinfo{year}{1969}), \bibinfo{pages}{576–580}.
\newblock
\showISSN{0001-0782}
\urldef\tempurl%
\url{https://doi.org/10.1145/363235.363259}
\showDOI{\tempurl}


\bibitem[Huang et~al\mbox{.}(2024)]%
        {huang2024enhancinglargelanguagemodels}
\bibfield{author}{\bibinfo{person}{Baizhou Huang}, \bibinfo{person}{Shuai Lu}, \bibinfo{person}{Weizhu Chen}, \bibinfo{person}{Xiaojun Wan}, {and} \bibinfo{person}{Nan Duan}.} \bibinfo{year}{2024}\natexlab{}.
\newblock \bibinfo{title}{Enhancing Large Language Models in Coding Through Multi-Perspective Self-Consistency}.
\newblock
\newblock
\showeprint[arxiv]{2309.17272}~[cs.CL]
\urldef\tempurl%
\url{https://arxiv.org/abs/2309.17272}
\showURL{%
\tempurl}


\bibitem[Jiang et~al\mbox{.}(2022)]%
        {jiang2022thor}
\bibfield{author}{\bibinfo{person}{Albert~Q. Jiang}, \bibinfo{person}{Wenda Li}, \bibinfo{person}{Szymon Tworkowski}, \bibinfo{person}{Konrad Czechowski}, \bibinfo{person}{Tomasz Odrzygóźdź}, \bibinfo{person}{Piotr Miłoś}, \bibinfo{person}{Yuhuai Wu}, {and} \bibinfo{person}{Mateja Jamnik}.} \bibinfo{year}{2022}\natexlab{}.
\newblock \bibinfo{title}{Thor: Wielding Hammers to Integrate Language Models and Automated Theorem Provers}.
\newblock
\newblock
\showeprint[arxiv]{2205.10893}~[cs.AI]


\bibitem[Kamath et~al\mbox{.}(2024)]%
        {kamath2024finding}
\bibfield{author}{\bibinfo{person}{Adharsh Kamath}, \bibinfo{person}{Aditya Senthilnathan}, \bibinfo{person}{Saikat Chakraborty}, \bibinfo{person}{Pantazis Deligiannis}, \bibinfo{person}{Shuvendu Lahiri}, \bibinfo{person}{Akash Lal}, \bibinfo{person}{Aseem Rastogi}, \bibinfo{person}{Subhajit Roy}, {and} \bibinfo{person}{Rahul Sharma}.} \bibinfo{year}{2024}\natexlab{}.
\newblock \bibinfo{booktitle}{\emph{Finding Inductive Loop Invariants using Large Language Models}}.
\newblock \bibinfo{type}{{T}echnical {R}eport} 2311.07948. \bibinfo{institution}{arXiv}.
\newblock


\bibitem[Lahiri(2024)]%
        {lahiri2024evaluatingllmdrivenuserintentformalization}
\bibfield{author}{\bibinfo{person}{Shuvendu~K. Lahiri}.} \bibinfo{year}{2024}\natexlab{}.
\newblock \bibinfo{title}{Evaluating LLM-driven User-Intent Formalization for Verification-Aware Languages}.
\newblock
\newblock
\showeprint[arxiv]{2406.09757}~[cs.PL]
\urldef\tempurl%
\url{https://arxiv.org/abs/2406.09757}
\showURL{%
\tempurl}


\bibitem[Liu et~al\mbox{.}(2024)]%
        {liu2024your}
\bibfield{author}{\bibinfo{person}{Jiawei Liu}, \bibinfo{person}{Chunqiu~Steven Xia}, \bibinfo{person}{Yuyao Wang}, {and} \bibinfo{person}{Lingming Zhang}.} \bibinfo{year}{2024}\natexlab{}.
\newblock \showarticletitle{{Is Your Code Generated by Chatgpt Really Correct? Rigorous Evaluation of Large Language Models for Code Generation}}.
\newblock \bibinfo{journal}{\emph{Advances in Neural Information Processing Systems}}  \bibinfo{volume}{36} (\bibinfo{year}{2024}).
\newblock


\bibitem[Mechtaev et~al\mbox{.}(2015)]%
        {directfix}
\bibfield{author}{\bibinfo{person}{Sergey Mechtaev}, \bibinfo{person}{Jooyong Yi}, {and} \bibinfo{person}{Abhik Roychoudhury}.} \bibinfo{year}{2015}\natexlab{}.
\newblock \showarticletitle{DirectFix: Looking for Simple Program Repairs}. In \bibinfo{booktitle}{\emph{2015 IEEE/ACM 37th IEEE International Conference on Software Engineering}}, Vol.~\bibinfo{volume}{1}. \bibinfo{address}{Florence, Italy}, \bibinfo{pages}{448--458}.
\newblock
\urldef\tempurl%
\url{https://doi.org/10.1109/ICSE.2015.63}
\showDOI{\tempurl}


\bibitem[Misu et~al\mbox{.}(2024)]%
        {misu2024towards}
\bibfield{author}{\bibinfo{person}{Md~Rakib~Hossain Misu}, \bibinfo{person}{Cristina~V. Lopes}, \bibinfo{person}{Iris Ma}, {and} \bibinfo{person}{James Noble}.} \bibinfo{year}{2024}\natexlab{}.
\newblock \showarticletitle{Towards AI-Assisted Synthesis of Verified Dafny Methods}.
\newblock \bibinfo{journal}{\emph{Proceedings of the ACM on Software Engineering}} \bibinfo{volume}{1}, \bibinfo{number}{FSE} (\bibinfo{date}{July} \bibinfo{year}{2024}), \bibinfo{pages}{812–835}.
\newblock
\showISSN{2994-970X}
\urldef\tempurl%
\url{https://doi.org/10.1145/3643763}
\showDOI{\tempurl}


\bibitem[Nashid et~al\mbox{.}(2023)]%
        {NashidICSE2023}
\bibfield{author}{\bibinfo{person}{Noor Nashid}, \bibinfo{person}{Mifta Sintaha}, {and} \bibinfo{person}{Ali Mesbah}.} \bibinfo{year}{2023}\natexlab{}.
\newblock \showarticletitle{{Retrieval-Based Prompt Selection for Code-Related Few-Shot Learning}}. In \bibinfo{booktitle}{\emph{{2023 IEEE/ACM 45th International Conference on Software Engineering (ICSE)}}}. \bibinfo{address}{Melbourne, Australia}, \bibinfo{pages}{2450--2462}.
\newblock
\urldef\tempurl%
\url{https://doi.org/10.1109/ICSE48619.2023.00205}
\showDOI{\tempurl}


\bibitem[Nguyen et~al\mbox{.}(2013)]%
        {semfix}
\bibfield{author}{\bibinfo{person}{Hoang Duong~Thien Nguyen}, \bibinfo{person}{Dawei Qi}, \bibinfo{person}{Abhik Roychoudhury}, {and} \bibinfo{person}{Satish Chandra}.} \bibinfo{year}{2013}\natexlab{}.
\newblock \showarticletitle{SemFix: Program repair via semantic analysis}. In \bibinfo{booktitle}{\emph{2013 35th International Conference on Software Engineering (ICSE)}}. \bibinfo{address}{San Francisco, CA, USA}, \bibinfo{pages}{772--781}.
\newblock
\urldef\tempurl%
\url{https://doi.org/10.1109/ICSE.2013.6606623}
\showDOI{\tempurl}


\bibitem[Ouyang et~al\mbox{.}(2023)]%
        {ouyang2023llm}
\bibfield{author}{\bibinfo{person}{Shuyin Ouyang}, \bibinfo{person}{Jie~M. Zhang}, \bibinfo{person}{Mark Harman}, {and} \bibinfo{person}{Meng Wang}.} \bibinfo{year}{2023}\natexlab{}.
\newblock \bibinfo{title}{LLM is Like a Box of Chocolates: the Non-determinism of ChatGPT in Code Generation}.
\newblock
\newblock
\showeprint[arxiv]{2308.02828}~[cs.SE]
\urldef\tempurl%
\url{https://arxiv.org/abs/2308.02828}
\showURL{%
\tempurl}


\bibitem[Pei et~al\mbox{.}(2023)]%
        {pmlr-v202-pei23a}
\bibfield{author}{\bibinfo{person}{Kexin Pei}, \bibinfo{person}{David Bieber}, \bibinfo{person}{Kensen Shi}, \bibinfo{person}{Charles Sutton}, {and} \bibinfo{person}{Pengcheng Yin}.} \bibinfo{year}{2023}\natexlab{}.
\newblock \showarticletitle{Can Large Language Models Reason about Program Invariants?}. In \bibinfo{booktitle}{\emph{Proceedings of the 40th International Conference on Machine Learning}} \emph{(\bibinfo{series}{Proceedings of Machine Learning Research}, Vol.~\bibinfo{volume}{202})}, \bibfield{editor}{\bibinfo{person}{Andreas Krause}, \bibinfo{person}{Emma Brunskill}, \bibinfo{person}{Kyunghyun Cho}, \bibinfo{person}{Barbara Engelhardt}, \bibinfo{person}{Sivan Sabato}, {and} \bibinfo{person}{Jonathan Scarlett}} (Eds.). \bibinfo{publisher}{PMLR}, \bibinfo{pages}{27496--27520}.
\newblock
\urldef\tempurl%
\url{https://proceedings.mlr.press/v202/pei23a.html}
\showURL{%
\tempurl}


\bibitem[Ringer(2021)]%
        {ringer2021proof}
\bibfield{author}{\bibinfo{person}{Talia Ringer}.} \bibinfo{year}{2021}\natexlab{}.
\newblock \bibinfo{booktitle}{\emph{Proof Repair}}.
\newblock \bibinfo{publisher}{University of Washington}, \bibinfo{address}{University of Washington}.
\newblock


\bibitem[Shoham et~al\mbox{.}(2007)]%
        {Sharon2007}
\bibfield{author}{\bibinfo{person}{Sharon Shoham}, \bibinfo{person}{Eran Yahav}, \bibinfo{person}{Stephen Fink}, {and} \bibinfo{person}{Marco Pistoia}.} \bibinfo{year}{2007}\natexlab{}.
\newblock \showarticletitle{Static specification mining using automata-based abstractions}. In \bibinfo{booktitle}{\emph{Proceedings of the 2007 International Symposium on Software Testing and Analysis}} (London, United Kingdom) \emph{(\bibinfo{series}{ISSTA '07})}. \bibinfo{publisher}{Association for Computing Machinery}, \bibinfo{address}{New York, NY, USA}, \bibinfo{pages}{174–184}.
\newblock
\showISBNx{9781595937346}
\urldef\tempurl%
\url{https://doi.org/10.1145/1273463.1273487}
\showDOI{\tempurl}


\bibitem[Sun et~al\mbox{.}(2024)]%
        {sun2023clover}
\bibfield{author}{\bibinfo{person}{Chuyue Sun}, \bibinfo{person}{Ying Sheng}, \bibinfo{person}{Oded Padon}, {and} \bibinfo{person}{Clark Barrett}.} \bibinfo{year}{2024}\natexlab{}.
\newblock \bibinfo{title}{Clover: Closed-Loop Verifiable Code Generation}.
\newblock
\newblock
\showeprint[arxiv]{2310.17807}~[cs.AI]
\urldef\tempurl%
\url{https://arxiv.org/abs/2310.17807}
\showURL{%
\tempurl}


\bibitem[{The dafny-lang community}(2024a)]%
        {dafnyfaq}
\bibfield{author}{\bibinfo{person}{{The dafny-lang community}}.} \bibinfo{year}{2024}\natexlab{a}.
\newblock \bibinfo{title}{{Dafny FAQ}}.
\newblock \bibinfo{howpublished}{\url{https://github.com/dafny-lang/dafny/wiki/FAQ\#how-does-dafny-handle-quantifiers-ive-heard-about-triggers-what-are-those}}.
\newblock


\bibitem[{The dafny-lang community}(2024b)]%
        {dafnyrefmanual}
\bibfield{author}{\bibinfo{person}{{The dafny-lang community}}.} \bibinfo{year}{2024}\natexlab{b}.
\newblock \bibinfo{title}{{Dafny Reference Manual}}.
\newblock \bibinfo{howpublished}{\url{https://dafny.org/dafny/DafnyRef/DafnyRef.html}}.
\newblock


\bibitem[Yang et~al\mbox{.}(2023)]%
        {yang2023leandojo}
\bibfield{author}{\bibinfo{person}{Kaiyu Yang}, \bibinfo{person}{Aidan~M. Swope}, \bibinfo{person}{Alex Gu}, \bibinfo{person}{Rahul Chalamala}, \bibinfo{person}{Peiyang Song}, \bibinfo{person}{Shixing Yu}, \bibinfo{person}{Saad Godil}, \bibinfo{person}{Ryan Prenger}, {and} \bibinfo{person}{Anima Anandkumar}.} \bibinfo{year}{2023}\natexlab{}.
\newblock \bibinfo{title}{LeanDojo: Theorem Proving with Retrieval-Augmented Language Models}.
\newblock
\newblock
\showeprint[arxiv]{2306.15626}~[cs.LG]
\urldef\tempurl%
\url{https://arxiv.org/abs/2306.15626}
\showURL{%
\tempurl}


\bibitem[Ziegler et~al\mbox{.}(2024)]%
        {CopilotImpact2024}
\bibfield{author}{\bibinfo{person}{Albert Ziegler}, \bibinfo{person}{Eirini Kalliamvakou}, \bibinfo{person}{X.~Alice Li}, \bibinfo{person}{Andrew Rice}, \bibinfo{person}{Devon Rifkin}, \bibinfo{person}{Shawn Simister}, \bibinfo{person}{Ganesh Sittampalam}, {and} \bibinfo{person}{Edward Aftandilian}.} \bibinfo{year}{2024}\natexlab{}.
\newblock \showarticletitle{{Measuring GitHub Copilot's Impact on Productivity}}.
\newblock \bibinfo{journal}{\emph{Commun. ACM}} \bibinfo{volume}{67}, \bibinfo{number}{3} (\bibinfo{date}{feb} \bibinfo{year}{2024}), \bibinfo{pages}{54–63}.
\newblock
\urldef\tempurl%
\url{https://doi.org/10.1145/3633453}
\showDOI{\tempurl}


\end{thebibliography}

\end{document}